\documentclass[journal]{IEEEtran}

\usepackage{cite}
\usepackage[dvips]{graphicx}
\DeclareGraphicsExtensions{.eps}
\usepackage[cmex10]{amsmath}
\usepackage{algorithmic}
\usepackage{algorithm}
\usepackage{amssymb}
\usepackage{psfrag}
\usepackage{graphicx}
\usepackage{cancel}
\usepackage{color}

\DeclareMathAlphabet{\mathbit}{OML}{cmr}{bx}{it}
\newcommand{\B}[1]{\mathbit{#1}}
\DeclareMathOperator{\Exp}{E}
\DeclareMathOperator{\Transpose}{T}
\DeclareMathOperator{\Hermitian}{H}
\DeclareMathOperator{\trace}{tr}
\DeclareMathOperator{\diag}{diag}
\DeclareMathOperator{\Fr}{F}

\newcommand{\Tr}{{\Transpose}}
\newcommand{\He}{{\Hermitian}}
\newcommand{\bs}{\boldsymbol}
\newcommand{\MMSE}{\text{MMSE}}
\newcommand{\MSE}{\text{MSE}}
\DeclareMathOperator*{\argmin}{argmin}

\ifCLASSOPTIONcompsoc \usepackage[caption=false,font=normalsize,labelfont=sf,textfont=sf]{subfig}
\else
 \usepackage[caption=false,font=footnotesize]{subfig}
\fi

\begin{document}
%
\title{QoS Constrained Power Minimization in the MISO Broadcast Channel with Imperfect CSI}

\author{Jos\'e P.~Gonz\'alez-Coma, 
        Michael~Joham,~\IEEEmembership{Member,~IEEE,}
        Paula M.~Castro,~\IEEEmembership{Member,~IEEE,}
        and~Luis~Castedo,~\IEEEmembership{Member,~IEEE,}
\thanks{This work has been supported by Xunta de Galicia, MINECO of Spain,
and FEDER funds of the EU under grants 2012/287 and  TEC2013-47141-C4-1-R.}
}

%



\maketitle

\begin{abstract}
We consider the design of linear precoders and receivers in a \emph{Multiple-Input Single-Output} (MISO) \emph{Broadcast Channel} (BC). We aim at minimizing the transmit power while fullfiling a set of per-user \emph{Quality-of-Service} (QoS) constraints expressed in terms of per-user average rate requirements. The \emph{Channel State Information} (CSI) is assumed to be perfectly known at the receivers but only partially at the transmitter. To solve the problem we transform the QoS constraints into \emph{Minimum Mean Square Error} (MMSE) constraints. We then leverage the MSE duality between the BC and the \emph{Multiple Access Channel} (MAC), as well as standard interference functions in the dual MAC, to perform power minimization by means of an \emph{Alternating Optimization} (AO) algorithm. Problem feasibility is also studied to determine whether the QoS constraints can be fulfilled or not. Finally, we present an algorithm to balance the average rates and manage situations that may be unfeasible or lead to an unacceptably high transmit power.
\end{abstract}



%
\IEEEpeerreviewmaketitle

\section{Introduction}
%
%
%
%
\IEEEPARstart{T}{he} \emph{Multiple-Input Single-Output} (MISO) \emph{Broadcast Channel} (BC) is an appropriate model for the downlink of a cellular communication system where a \emph{Base Station} (BS) with $N$ antennas serves a set of $K$ single-antenna non-cooperative users. We assume signals are linearly filtered at transmission and reception to mitigate the inter-user interference. We also assume perfect \emph{Channel State Information at the Receivers} (CSIR) but only imperfect \emph{Channel State Information at the Transmitter}  (CSIT). This is a reasonable assumption in practical setups since receivers can accurately estimate the CSI from the incoming signals whereas the transmitter obtains the CSI via a feedback channel in \emph{Frequency Division Duplex} (FDD) systems, or an estimate of the reciprocal uplink CSI in \emph{Time Division Duplex} (TDD) systems.

Several imperfect CSI models have been considered in the literature. Some authors employ bounded uncertainty models such as  ellipsoidal \cite{MuKiBo07}, spherical \cite{BiYoHa12,ShLa08,VuBo08,VuBo09}, or rectangular \cite{VuBo09}, and formulate worst-case performance optimization problems that can be solved using \emph{Semi-Definite Program} (SDP) methods \cite{BoVa04}. Other authors, as done in this work, model CSI uncertainty as a stochastic error whose distribution is known in single-user \cite{ViMa01,JoBo04} and multiple-user \cite{ShDa08,ShDa08Asilomar,UbCh08,BoVa12,KoCa07,CaJiKoRa10,NeGhSl12,RaBoZh13,SoUl07,JoCl14} scenarios. 

Different performance metrics have been considered for the BC optimization.   Maximizing the \emph{Signal to Interference--plus--Noise Ratio} (SINR) \cite{ScBo07,VuBo08,VuBo09,MuKiBo07,BiYoHa12,ShLa08,ChLi07,ShLuCh08,BoHuTr07,LiZhLiYa11,WuWaCa08,ShDa07,ShDa08Asilomar}, is a common approach closely related to the maximization of the data rate. Moreover, in \cite{ScBo04,ScBo07,KoCa07,ShLa08,ChLi07} imperfect CSIT is considered by handling approximations for the average SINR where the expectation is separately applied to the numerator and the denominator. The tightness of such an approximation, however, is questionable and it is unclear whether the approximation is an upper or a lower bound. Other metrics are based on the \emph{Mean Square Error} (MSE). Per-user MSE was considered in \cite{ShScBo08,ShDa09,VuBo09} or recently in \cite{JoCl14}, where an approximation of the average MSE based on a Taylor expansion has been proposed. Sum MSE  \cite{ShDa08,UbCh08,BoVa12,BoChVa11,ShScBo07}, and MSE balancing \cite{ShScBo08,ShDa08,BoChVa11} have also been often addressed. The sum MSE minimization in the BC can be transformed into an equivalent one in the dual \emph{Multiple Access Channel} (MAC) to perform \emph{Alternate Optimization} (AO). 
Finally, weighted sum rate was studied in \cite{KoCa07,ShScBo08TSP,NeGhSl12,ChAgCaCi08}. A common approach is to reformulate the problem as a weighted sum MSE to find solutions based on \emph{Geometric Programing} (GP), or on the algorithm proposed in \cite{ChAgCaCi08}. However, sum rate optimizations may lead to unfair and non-desirable situations where some of the users get low (or even zero) information rates.

Regarding the optimization in the BC, some authors search for the best metric performance for given transmit power \cite{ShDa08,VuBoSh09,UbCh08,BoVa12,BiYoHa12,ChLi07,BoHuTr07,WuWaCa08,BoChVa11,NeGhSl12,ShDa07}. Contrary to that, authors in \cite{VuBo09,VuBoSh09,VuBo08,MuKiBo07,ShLa08,KoCa07,ShDa07,ShDa09,ShDa08Asilomar} consider the minimization of the total transmit power under a set of \emph{Quality-of-Service} (QoS) constraints, as done in this work. In particular, we ensure that users enjoy certain average rate values. Note that such restrictions make it possible to avoid the  unfair situations stated previously. 

To tackle this optimization problem, average rate constraints are replaced by average MMSE requirements using Jensen's inequality (see also \cite{StYiJi05}). Note that, contrary to other solutions (e.g. \cite{ScBo04,JoCl14}), no approximations are needed to theoretically solve the MSE problem formulation. Hence, we determine the MISO BC linear precoders and receivers by means of an AO process in which we resort to the duality between the BC and the MAC, as done in, e.g., \cite{ShScBo07,BoVa12}, to design the transmit and receive filters. More specifically, we employ the MSE duality proposed in \cite{JoVoUt10} for the assumptions of perfect CSIR and imperfect CSIT.

In the dual MAC, power minimization can be formulated as a power allocation problem and solved using the standard interference function framework proposed in \cite{Yates} and extended in \cite{BoSc07}. 

This work also shows that the proposed power minimization algorithm converges if the QoS constraints can be fulfilled. 
Therefore, we provide a test for checking the feasibility of the average rate restrictions. This test is a generalization of that presented in \cite{HuJo10} for the vector BC and perfect CSIT and CSIR.

Additionally, we consider the rate balancing problem: the minimum of the average rates is maximized under a total transmit power constraint. Again, this problem is reformulated bounding the average rates by average MMSEs. Such a reformulation leads to the minimization of the maximum weighted average MSE under a total power constraint, and can be solved combining a bisection search with the proposed power minimization algorithm.

In recent communication systems, users are equipped with more than one antenna. When we extend the system model to the MIMO scenario two directions arise: considering single and multiple per-user streams. Considering more than one per-user stream adds more complexity to the problem, since the per-user average rate constraints have to be divided between all the streams allocated to the user. Such discussion is out of the scope of this work. However, the methods proposed for the MISO BC directly apply in the single-stream MIMO BC, as shown in \cite{Our2}.

The paper is organized as follows. Section \ref{sec:sysmodel} describes the MISO BC system model and the BC/MAC MSE duality. Section \ref{sec:probForm} addresses the power minimization problem using the standard interference function framework and an AO approach. Section \ref{sec:feasibility} considers the feasibility of the QoS constraints while Section \ref{sec:balancing} considers the rate balancing problem. Finally, the results of simulation experiments are given in Section \ref{sec:simul} and the conclusions in Section \ref{conclusions}.  

The following notation is employed. Matrices and column vectors are written using upper an lower boldface characters, respectively. By $[\B{X}]_{j,k}$, we denote the element in row $j$ and column $k$ of the matrix $\B{X}$; $\diag(x_i)$ represents a diagonal matrix whose $i$th diagonal element is $x_i$; $\mathbf{I}_N$ stands for the $N\times N$ identity matrix, and $\mathbf{1}$ represents the all ones vector. The superscripts $(\cdot)^*$, $(\cdot)^\Tr$, and $(\cdot)^\He$ denote the complex conjugate, transpose, and Hermitian. $\Re\{\cdot\}$ represents the real part operator.  Finally, $\Exp[\cdot]$ stands for statistical expectation, $\trace(\cdot)$ denotes the trace operation, and $|\cdot|$, $\|\cdot\|_2$, $\|\cdot\|_{\Fr}$ stand for the absolute value, the Euclidean norm, and the Frobenius norm, respectively.


\section {System Model}
\label{sec:sysmodel}
\begin{figure}[t]
  	\psfrag{ss1}{$s_1$}
  	\psfrag{ss2}{$s_1$} 
	\psfrag{ssk}{$s_K$} 
	\psfrag{p1}{$\B{p}_1$}
	\psfrag{p2}{$\B{p}_2$}
	\psfrag{pk}{$\B{p}_K$}
	\psfrag{h1}{$\B{h}_1^{\He}$}
	\psfrag{h2}{$\B{h}_2^{\He}$}
	\psfrag{hk}{$\B{h}_K^{\He}$}
	\psfrag{n1}{$\eta_1$}
	\psfrag{n2}{$\eta_2$}
	\psfrag{nk}{$\eta_K$}
	\psfrag{y}{}
	\psfrag{x1}{}
	\psfrag{xK}{}
	\psfrag{sh1}{$\hat{s}_1$}
	\psfrag{sh2}{$\hat{s}_2$}
	\psfrag{shk}{$\hat{s}_K$}
	\psfrag{f1}{$f_1$}
	\psfrag{f2}{$f_2$}	
	\psfrag{fk}{$f_K$}
	\includegraphics[width=\linewidth]{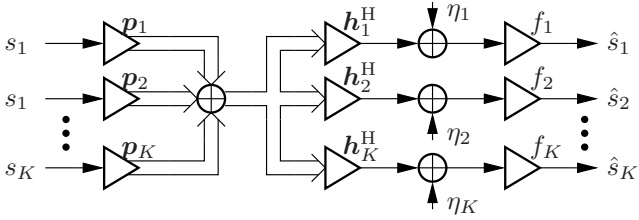}
	\caption{Sytem model of the Gaussian MISO BC.}
\label{fig:bcsystem}
\end{figure}

Let us consider the system model of a Gaussian MISO BC depicted in Fig. \ref{fig:bcsystem}. We assume the BS is equipped with $N$ transmit antennas and sends the data signal $s_k\in \mathbb{C}$ to the user $k\in\{1,\ldots,K\}$. The data signal vector ${\bf s} = [s_1, \ldots, s_K]^T$ is assumed to be zero-mean, unit-variance, uncorrelated, and Gaussian, i.e., ${\bf s} \sim \mathcal{N}_{\mathbb{C}}(\mathbf{0},\mathbf{I}_K$). The data signals are precoded with the linear filters $\B{p}_k\in \mathbb{C}^{N}$ at the BS and propagate over the vector channels $\B{h}_k\in \mathbb{C}^{N}$. At the users-ends, the received signals are linearly filtered with $f_k\in\mathbb{C}$ to produce an estimate of the $k$-th user data signal
\begin{equation}
\hat{s}_k=f_k^*\B{h}_k^\He\sum_{i=1}^{K}\B{p}_i s_i+f_k^*\eta_k,
\label{hatsk}
\end{equation}
where $\eta_k\sim\mathcal{N}_{\mathbb{C}}(0,\sigma^2_{\eta_k})$ represents the thermal noise which is independent of the data signals. Note that, according to this signal model the transmit power is $\sum_{k=1}^{K}\left\|\B{p}_k\right\|^2$.

We assume that the receiver $k$ has perfect knowledge of its own channel $\B{h}_k$. Contrarily, the BS has only imperfect knowledge of the CSI which is  modeled through the random variable $v$. The random nature of $v$ is due to numerous sources of error (i.e., channel estimation, quantization, delay, \ldots) that affect the process of acquiring the CSIT in both TDD and FDD systems. The imperfect channel knowledge is expressed through the conditional \emph{Probability Density Functions} (PDF) $f_{\B{h}_k|v}(\B{h}_k|v)$, assumed to be known at the transmitter.

Recalling (\ref{hatsk}), $\hat{s}_k$ is a noisy version of the data signal $s_k$. The achievable instantaneous data rate in such situation is
\begin{equation}
R_k=\log_2(1+\B{p}_k^\He\B{h}_k \B{h}_k^\He\B{p}_kx_k^{-1}),
\label{eq:rate}
\end{equation}
where $x_k=\B{h}_k^\He(\sum_{i\neq k}\B{p}_i\B{p}_i^\He)\B{h}_k+\sigma^2_{\eta_k}$. In this work we search for the precoders $\B{p}_k$ that minimize the transmit power fulfilling the Quality of Service (QoS) constraints $\Exp[R_k(v)]\geq\rho_k$, $k\in\{1,\ldots,K\}$, where $\{\rho_k\}_{k=1}^K$ is the set of per-user average rates to be fulfilled by the system. Note that the notation $R_k(v)$ highlights that the transmitter has access to the partial CSIT $v$ for any channel realization $\B{h}_k,\,\forall  k$. Based on partial CSIT $v$, the BC precoders are determined according to the variational problem
\begin{equation}
\min_{\{\B{p}_k(v)\}_{k=1}^K} \Exp\left[\sum_{k=1}^{K}\left\|\B{p}_k(v)\right\|^2_2\right]\quad\text{s.t.}\quad\Exp\left[R_k(v)\right]\geq\rho_k,\,\forall k.
\label{eq:powermin}
\end{equation}

Note that the optimization is over the maps $\B{p}_k(v)$, i.e., the precoders depending on the partial CSIT $v$. The constrained minimization problem \eqref{eq:powermin} is difficult to solve in general. However, in the ensuing subsection, we exploit the relationship between the average rate and the average MMSE to reformulate (\ref{eq:powermin}) in a more manageable way.

\subsection{MSE Constrained Optimization}

Let $\text{MSE}_k^{\text{BC}}=\Exp[|s_k-\hat{s}_k|^2]$ be the instantaneous MSE of the $k$-th user in the BC. For given channel $\B{h}_k$,
\begin{align}
\text{MSE}_k^{\text{BC}} & = 1-2\Re\left\lbrace f_k^*\B{h}_k^\He\B{p}_k\right\rbrace+\left|f_k\right|^2\left(\left|\B{h}_k^\He\B{p}_k\right|^2+x_k\right), \label{eq:MSEBC}
\end{align}
where $x_k$ is that defined below \eqref{eq:rate}. Note that $\B{h}_k$ is assumed to be fixed in \eqref{eq:MSEBC}. Therefore, also the partial CSIT $v$ is fixed and we drop the dependence of $\B{p}_k$ on $v$ for the sake of brevity. Correspondingly, the minimum MSE receive filter is given by
\begin{equation}
f_k^\MMSE(\B{h}_k)=\left(\B{h}_k^\He\sum_{i=1}^{K}\B{p}_i\B{p}_i^\He\B{h}_k+\sigma^2_{\eta_k}\right)^{-1}\B{h}_k^\He\B{p}_k,
\label{eq:Fmmse}
\end{equation}
and the MMSE is obtained substituting \eqref{eq:Fmmse} into \eqref{eq:MSEBC}, i.e.,
\begin{equation}
\MMSE_k^\text{BC}=1-f_k^{\MMSE,*}(\B{h}_k)\B{h}_k^\He\B{p}_k.
\label{eq:MMSEBC}
\end{equation}
Finally, by applying the equality $1-\frac{a}{b}=(1+\frac{a}{b-a})^{-1}$ to \eqref{eq:MMSEBC} it is possible to express the $k$-th user rate \eqref{eq:rate} as $R_k=-\log_2(\MMSE_k^\text{BC})$ (cf. \cite{StYiJi05}). 

Equations \eqref{eq:MSEBC}, \eqref{eq:Fmmse} and \eqref{eq:MMSEBC} are suitable for the BC design with perfect CSI at both ends of the communication system. Notice, however, that imperfect CSIT is assumed in this work. For this reason, consider the average MSE at the BC $\Exp[\text{MSE}_k^{\text{BC}}(v)]$. Correspondingly, the average MMSE at the BC is given by 
\begin{equation}
\Exp[\text{MMSE}_k^{\text{BC}}(v)] = \Exp\left[1-f_k^{\MMSE,*}\left(\B{h}_k\right)\B{h}_k^\He\B{p}_k(v)\right],
\nonumber
\end{equation}
where we highlight the perfect CSIR assumption by $f_k(\B{h}_k)$.

Taking advantage of the concavity of the $\log_2(\cdot)$ function and employing Jensen's inequality, we arrive at the following lower bound for the average rate 
\begin{equation}
\Exp\left[R_k(v)\right]\geq-\log_2\Exp\left[\MMSE_k^\text{BC}(v)\right]\geq -\log_2\Exp\left[\MSE_k^\text{BC}(v)\right]
\label{eq:MMSElb}
\end{equation}
An example of the gap between the average rate and the average MMSE lower bound is examined in Appendix \ref{sec:gap}.

The constraints in \eqref{eq:powermin} hold for $-\log_2\Exp[{\MSE}_k^{\text{BC}}(v)] \geq \rho_k$, and they are conservatively rewritten accordingly as 
\begin{equation}
\Exp\left[\MSE_k^\text{BC}(v)\right]\leq 2^{-\rho_k}.
\label{eq:MMSEconstr}	
\end{equation} 
Hence, the optimization problem  \eqref{eq:powermin} can be reformulated as 
\begin{align}
\min_{\{\B{p}_k(v),f_k(\B{h}_k)\}_{k=1}^K}&\Exp\left[\sum_{k=1}^{K}\left\|\B{p}_k\left(v\right)\right\|^2_2\right]\nonumber\\
&\text{s.t.}\quad\Exp\left[\MSE_k^{\text{BC}}(v)\right]\leq 2^{-\rho_k},\,\forall k.\label{eq:BCreform1}
\end{align}
Contrary to \eqref{eq:powermin}, the scalar receive filters $f_k(\B{h}_k)$ are now involved in the optimization process. Nevertheless, in the optimum of \eqref{eq:BCreform1}, MMSE filters are employed [see \eqref{eq:Fmmse}]. 

We now note that by means of Bayes' rule $\Exp[\MSE^{\text{BC}}_k(v)]=\Exp[\Exp[\MSE^{\text{BC}}_k(v)|\,v]]$. 
Then, introducing $\overline{\MSE}_k^{\text{BC}}(v)=\Exp[\MSE^{\text{BC}}_k(v)|\,v]$, the variational problem \eqref{eq:BCreform1} can be solved pointwise for given $v$ as follows
\begin{equation}
\min_{\{\B{p}_k(v),f_k(\B{h}_k)\}_{k=1}^K}\sum_{k=1}^{K}\left\|\B{p}_k(v)\right\|^2_2\,\text{s.t.}\,\,\overline{\MSE}_k^{\text{BC}}\left(v\right)\leq 2^{-\rho_k},\,\forall k.
\label{eq:BCreform}
\end{equation}
Note that the average transmit power resulting from \eqref{eq:BCreform} is larger than that obtained from \eqref{eq:powermin} since the MMSE constraints in \eqref{eq:BCreform} are more restrictive than the rate constraints in \eqref{eq:powermin}.In the following, we use $\B{p}_k$, $f_k$ and $\overline{\MSE}_k^{\text{BC}}$ for the sake of notational brevity.


\subsection{BC/MAC MSE Duality}
\label{sec:duality}

It is important to note that $\overline{\text{MSE}}_k^{\text{BC}}$ is independent of the receive filter $f_j$ for $j \neq k$ but depends on all precoders $\B{p}_j$ for $j \neq k$. This means that $\B{p}_k$ cannot be individually optimized when solving \eqref{eq:BCreform} but all precoders should be optimized jointly. Nevertheless, it is possible to avoid such dependence by exploiting the MAC/BC MSE duality described in \cite{JoVoUt10}.

In the \emph{Single-Input Multiple-Output} (SIMO) MAC dual to the MISO BC, the receive and transmit filters are represented by $\B{g}_k\in \mathbb{C}^{N}$ and $t_k\in \mathbb{C}$, respectively, while  $\bs{\theta}_k=\B{h}_k\sigma_{\eta_k}^{-1}\in \mathbb{C}^{N}$ and $\B{n}\sim\mathcal{N}_{\mathbb{C}}(\mathbf{0},\mathbf{I}_N)$ represent the channel response and noise in the dual MAC, respectively. The average MSE is then
\begin{align}
\overline{\text{MSE}}_{k}^{\text{MAC}}(v)&=1 - 2 \Exp\left[\Re\left\lbrace\B{g}^\He_{k}\bs{\theta}_kt_{k}\right\rbrace|\,v\right]+\left\|\B{g}_{k}\right\|^2_2\nonumber\\
&+\Exp\left[\left.\sum_{i=1}^K\left|t_{i}\right|^2\left|\B{g}^\He_{k}\bs{\theta}_i\right|^2\right|\,v\right],
\label{eq:avgMACMSE}
\end{align}
where the expectations are taken w.r.t. all channels for given partial CSI $v$ as in $\overline{\text{MSE}}_{k}^{\text{BC}}(v)$ from \eqref{eq:BCreform}.

Suppose now that the filters in the MAC, i.e., $t_k$ and $\B{g}_k$, are given. Introducing the set  $\{\alpha_k\}_{k=1}^{K}\in\mathbb{R}^{+}$, and the following relationships between the MAC and the BC filters 
\begin{align}
\B{p}_{k}(v) & = \alpha_{k}\B{g}_{k}(v),  \nonumber \\
f_{k} & = \alpha_{k}^{-1}\sigma_{\eta_k}^{-1}t_{k}\left(\bs{\theta}_1,\bs{\theta}_2,\ldots,\bs{\theta}_K\right),
\label{eq:filters_rel}
\end{align}
it is possible to achieve identical MSEs for all the users in the BC as in the MAC, i.e., $\overline{\text{MSE}}_{k}^{\text{BC}}=\overline{\text{MSE}}_{k}^{\text{MAC}}\,\forall k$. Moreover, the average transmit power is preserved \cite{JoVoUt10}. Note that even not always explicitly remarked in the notation, the MAC receive filters and precoders are functions of the partial CSIT $v$ and the channel, respectively, as the corresponding BC precoders and receive filters.   

In summary, a problem in the BC based on $\overline{\text{MSE}}_k^\text{BC}$ can be equivalently reformulated in the dual MAC with $\overline{\text{MSE}}_k^\text{MAC}$, and vice-versa. This duality result will be exploited in the ensuing sections to determine the BC precoders $\B{p}_k$.

\section{Power Minimization}
\label{sec:probForm}

We now focus on solving the power minimization problem as formulated in \eqref{eq:BCreform}. 
First of all, for given BC precoders $\B{p}_k$, the MMSE BC scalar receive filters $f_k^\MMSE$ are readily obtained via \eqref{eq:Fmmse} considering perfect CSIR. Next, we transform the BC receive filters $f_k$ into the MAC precoding weights $t_k$ using the MSE duality. Recall that $t_k$ is a function of $\B{h}_k$.

Let us now define the average transmit power $\xi_k=\Exp[|t_k|^2|\,v]$ and the normalized MAC precoders $\tau_k=t_k/\sqrt{\xi_k}$ such that $\Exp[|\tau_k|^2|\,v]=1$. Let us also introduce the conditional expectations $\bs{\mu}_k=\Exp[\tau_k\bs{\theta}_k|\,v]$ and $\B{\Theta}_i=\Exp[|\tau_i|^2\bs{\theta}_i\bs{\theta}_i^\He|\,v]$. Finally, let us define $\bs{\xi}=[\xi_1,\ldots,\xi_K]^{\Tr}$ as the vector containing the average transmit powers for all users, i.e., the power allocation vector. Notice that, unlike the precoders $t_k$, $\bs{\xi}$ only depends on the partial CSIT $v$, similar to the total transmit power $\sum_{k=}^{K}\|\B{p}_k(v)\|_2^2$ in the BC.

With these definitions, $\overline{\text{MSE}}_{k}^{\text{MAC}}$ from \eqref{eq:avgMACMSE} reads as 
\begin{align}
\overline{\text{MSE}}_{k}^{\text{MAC}}= 1-2\sqrt{\xi_k}\Re\left\lbrace\B{g}^\He_{k}\bs{\mu}_{k}\right\rbrace+\B{g}_{k}^{\He}\left(\sum_{i=1}^K\xi_i\B{\Theta}_i+\mathbf{I}_N\right)\B{g}_{k}.
\label{eq:ExpMMSE}
\end{align} 
Therefore, the equalizers minimizing the $\overline{\text{MSE}}_{k}^{\text{MAC}}$ are
\begin{equation}
\B{g}_{k}^{\MMSE} = \left(\sum_{i=1}^{K}\xi_i\B{\Theta}_i+\mathbf{I}_N\right)^{-1}\sqrt{\xi_k}\bs{\mu}_k.
\label{eq:gmmse}
\end{equation}
By substituting \eqref{eq:gmmse} into \eqref{eq:ExpMMSE}, we obtain the following expression for the average MMSE conditioned on $v$
\begin{align}
\overline{\MMSE}^{\text{MAC}}_{k}&=1-\xi_k\bs{\mu}_k^{\He}\left(\sum_{i=1}^{K}\xi_i\B{\Theta}_i+\B{I}_N\right)^{-1}\bs{\mu}_k.
\label{eq:MISOMACavgMMSE}
\end{align}

We now show that a scaled version of $\B{g}_{k}^{\MMSE}$ also minimizes the $\overline{\text{MSE}}_{k}^{\text{MAC}}$ given by \eqref{eq:ExpMMSE}. This result will be exploited later on to obtain a simple update of the equalizers in the iterative algorithm that minimizes the transmit power. Let us introduce the scalar MAC parameters $r_k$ so that $\B{g}_k=r_k\tilde{\B{g}}_k$. With this new notation, the $\overline{\text{MSE}}_{k}^{\text{MAC}}$ in \eqref{eq:ExpMMSE} reads as
\begin{align}
\overline{\text{MSE}}_{k}^{\text{MAC}}&=1-2\Re\left\lbrace r_k^*\tilde{\B{g}}^\He_{k}\bs{\mu}_{k}\sqrt{\xi_k}\right\rbrace\nonumber\\
&+\left|r_k\right|^2\tilde{\B{g}}_{k}^{\He}\left(\sum_{i=1}^K\xi_i\B{\Theta}_i+\mathbf{I}_N\right)\tilde{\B{g}}_{k}.
\label{eq:MACavgMSE}
\end{align}
For given $\tilde{\B{g}}_k$, the optimal scalar filters are
\begin{equation}
r_k^\MMSE=\tilde{\B{g}}^\He_{k}\bs{\mu}_k\sqrt{\xi_k}\left(\tilde{\B{g}}_{k}^\He\left(\sum_{i=1}^K\xi_i\B{\Theta}_i+\mathbf{I}_N\right)\tilde{\B{g}}_{k}\right)^{-1}.
\end{equation}
Substituting $r_k^\MMSE$ into (\ref{eq:MACavgMSE}) yields the following minimum average MAC MSE
\begin{equation}
\Sigma_{k}=1-\xi_k\left|\tilde{\B{g}}^\He_{k}\bs{\mu}_{k}\right|^2y_k^{-1},
\label{eq:MACscalar}
\end{equation}where $y_k=\tilde{\B{g}}_{k}^\He(\sum_{i=1}^K\xi_i\B{\Theta}_i+\mathbf{I}_N)\tilde{\B{g}}_{k}$.
Note now that replacing $\tilde{\B{g}}_k$ in \eqref{eq:MACscalar} by $\B{g}_{k}^{\MMSE}$ given by \eqref{eq:gmmse},  leads to \eqref{eq:MISOMACavgMMSE}. Therefore, \eqref{eq:gmmse} is the minimizer of  \eqref{eq:ExpMMSE} and \eqref{eq:MACscalar}. 


\subsection{Power Allocation}

So far, we have found the MMSE vector receivers in the MAC, $\{\B{g}_{k}^{\MMSE}\}_{k=1}^K$, corresponding to the BC precoders $\{\B{p}_k\}_{k=1}^K$. We now search for the optimal MAC receivers $\{\B{g}_k\}_{k=1}^K$ and power allocation $\bs{\xi}$ that minimize the transmit power (subject to the QoS constraints $\overline{\MMSE}_k^{\text{BC}}\leq 2^{-\rho_k}$) for given normalized precoders $\{\tau_k\}_{k=1}^K$. Due to the mutual dependence of $\{\B{g}_k\}_{k=1}^K$ and $\bs{\xi}$, we have to jointly optimize both of them. 

To that end, we rely on standard interference functions \cite{Yates, ScBo07}. Interference functions concisely describe the framework of the system requirements depending on the power allocation as the vector inequality $\bs{\xi}\geq\B{f}(\bs{\xi})$. To ensure that the fixed point iteration $\bs{\xi}^{(n+1)}=\B{f}(\bs{\xi}^{(n)})$ converges to the optimal solution for $\bs{\xi}$, the function $\B{f}(\cdot)$ must be a standard interference function, i.e., it satisfies
\begin{itemize}
\item[] $\B{f}(\bs{\xi})>\mathbf{0}$ (positivity)
\item[] $a\B{f}(\bs{\xi})>\B{f}(a\bs{\xi})$ $\forall a>1$ (scalability), and
\item[] $\B{f}(\bs{\xi})\geq \B{f}(\bs{\xi}'),\,\bs{\xi}\geq\bs{\xi}'$ ( monotonicity).
\end{itemize}

We now define $I_k(\bs{\xi})=\xi_k \Sigma_k$ which can be interpreted as the interference for user $k$. Applying the equality $1-\frac{a}{b}=(1+\frac{a}{b-a})^{-1}$ to \eqref{eq:MACscalar} gives 
\begin{equation}
I_k\left(\bs{\xi}\right)=\left(\frac{1}{\xi_k}+\left|\tilde{\B{g}}^\He_{k}\bs{\mu}_{k}\right|^2\left(y_k-\xi_k\left|\tilde{\B{g}}^\He_{k}\bs{\mu}_{k}\right|^2\right)^{-1}\right)^{-1}.
\label{eq:interf}
\end{equation}
We next collect all these functions into the vector $\B{I}(\bs{\xi})=[I_1(\bs{\xi}),\ldots,I_K(\bs{\xi})]$. As shown in Appendix \ref{ap:intfunc}, $\B{I}(\bs{\xi})$ fulfills the properties of a standard interference function.

Note that, due to the average MSE BC/MAC duality, the QoS constraints can equivalently be expressed as $\overline{\MSE}_k^\text{MAC}\leq 2^{-\rho_k}$. Furthermore, since  $\Sigma_k=\frac{I_k(\bs{\xi})}{\xi_k}$, we reformulate the power minimization problem \eqref{eq:BCreform} in the dual MAC for a given set of normalized precoders $\{\tau_k\}_{k=1}^K$ as
\begin{equation}
\min_{\{\xi_k,\tilde{\B{g}}_k\}_{k=1}^K}\sum_{i=1}^{K}\xi_i\quad\text{s.t.} \quad \frac{I_k\left(\bs{\xi}\right)}{\xi_k} \leq 2^{-\rho_k},\,\forall k.
\label{eq:macform}
\end{equation}  

As shown in \cite{Yates}, since $\B{I}(\bs{\xi})$ is a standard interference function, the iteration $\xi_k^{(n)}=2^{\rho_k}I_k(\bs{\xi}^{(n-1)})$ converges to $\xi_k^\text{opt}$ for given $\{\tilde{\B{g}}_k\}_{k=1}^K$. 

Moreover, the previously mentioned iteration can also be used to jointly find the $\{\xi_k,\tilde{\B{g}}_k\}_{k=1}^K$ that solve the power minimization problem \eqref{eq:macform}. Indeed, let $I_k(\bs{\xi},\tilde{\B{g}}_k)=\xi_k\Sigma_k$ be the same function as before, but explicitly highlighting the dependence on $\tilde{\B{g}}_k$. Similarly, we rewrite the interference function as $\B{I}(\bs{\xi},\tilde{\B{G}})=[I_1(\bs{\xi},\tilde{\B{g}}_1),\ldots,I_K(\bs{\xi},\tilde{\B{g}}_K)]^\Tr$ with $\tilde{\B{G}}=[\tilde{\B{g}}_1,\ldots,\tilde{\B{g}}_K]$. Since $\B{I}(\bs{\xi},\tilde{\B{G}})$ is standard for any $\tilde{\B{G}}$, so is $\min_{\tilde{\B{G}}}\B{I}(\bs{\xi},\tilde{\B{G}})$ where the minimization is performed element-wise. As a consequence, the Alternating Optimization (AO) iteration 
\begin{align}
\tilde{\B{g}}_k^{(n)} & \leftarrow \argmin_{\tilde{\B{g}}_k}I_k\left(\bs{\xi}^{(n-1)},\tilde{\B{g}}_k\right) \forall k, \nonumber\\
\xi_k^{(n)} & \leftarrow 2^{\rho_k} I_k\left(\bs{\xi}^{(n-1)},\tilde{\B{g}}_k^{(n)}\right) \forall k,
\label{eq:iteration}
\end{align}
converges to the global optimum of \eqref{eq:macform}, as shown in \cite{ScBo07}.

Finally, the obtained dual MAC equalizers can be transformed into the BC precoders by applying the average MSE BC/MAC duality [see \eqref{eq:filters_rel}]. Afterwards, the BC MMSE receive filters can be updated for these BC precoders. The iterative process that alternates between the optimization of both filters is referred to as AO.

\subsection{Power Minimization Algorithm}
\label{sec:algorithm1}
Algorithm \ref{alg:pwrmin} presents the steps to solve the optimization problem \eqref{eq:BCreform} according to the ideas presented so far. 

Recall that we assume $v$ and  $f_{\B{h}_k|v}(\B{h}_k|v)$ are known at the transmitter according to th imperfect CSIT model. Since closed-form expressions of the expectations in \eqref{eq:macform} are not known for general channel models, we evaluate them by using a Monte Carlo method. To that end, we generate $M$ channel realizations $\B{h}_k^{(m)}\sim f_{\B{h}_k|v}(\B{h}_k|v)$, $m=1, \ldots , M$, and introduce the matrix $\B{H}_k=\sigma_{\eta_k}^{-1}[\B{h}_k^{(1)},\ldots,\B{h}_k^{(M)}]$ to collect the $M$ dual MAC channel realizations.
We also define $t_k^{(m)}$ as the $k$-th user scalar MAC precoder for given channel realization $\B{h}_k^{(m)}$. Collecting the $t_k^{(m)}$ we get the normalized precoding diagonal  matrix
\begin{equation}
\B{T}_k=\frac{1}{\sqrt{\xi_k}} \diag\left(t_k^{(1)},\ldots,t_k^{(M)}\right),
\end{equation} 
where $\xi_k=\frac{1}{M}\sum_{m=1}^{M}|t_k^{(m)}|^2$ is the $k$-th user average transmit power for given $v$. Therefore, we calculate the expectations as $\bs{\mu}_k=\frac{1}{M}\B{H}_k\B{T}_k\mathbf{1}$ and $\B{\Theta}_k=\frac{1}{M}\B{H}_k\B{T}_k\B{T}_k^\He\B{H}_k^\He$. 

We start with an initial set of BC random precoders $\{\B{p}_k^{(0)}\}_{k=1}^{K}$ (line 1). We next calculate the $M$ BC receivers $f_k^{\MMSE,(m)}$ corresponding to the channel realizations $\B{h}_k^{(m)}$ (line 5). Applying the BC/MAC duality we determine the $M$ dual MAC precoders (line 7). The normalized matrix of MAC precoders is obtained after the execution of lines 8 and 9.  

The following two steps (lines 10 and 11) perform iteration \eqref{eq:iteration} to update the power allocation and the dual MAC receivers. Observe, however, that we do not include the loop arising from the optimization in \eqref{eq:iteration}. The reason is to avoid convergence problems, which may occur even when the problem constraints are feasible,  caused by the non-feasibility of the power minimization problem for given MAC precoders $\B{T}_k^{(\ell)}$ at the $\ell$-th iteration (cf. \eqref{eq:macform}). Therefore, considering a single loop we avoid this undesirable effect, as can be appreciated from our simulation experiments (cf.\cite{Our,Our2,Our3}). 

After the power allocation and the receive filters update (lines 10 and 11), the new MAC transmit filters are determined in line 13. Finally, we switch back to the BC in line 15. Due to the existence of a unique minimum in \eqref{eq:BCreform}, and to the fact that every step in the algorithm either reduces the average MMSEs or the total transmit power, the convergence of the algorithm is guaranteed when the QoS constraints are feasible (see Section \ref{sec:feasibility}). To check whether we have reached the desired accuracy or not, we set a threshold $\delta$ (line 16). 

Note that the algorithm computational complexity is approximately linear in the number of channel realizations, $\mathcal{O}(M)$, since the sizes of the matrices to be inverted in lines 7, 15 and 11 are small compared to $M$, i.e. $K\ll M$ and $N\ll M$.
\begin{algorithm}[t]
\caption{Power Minimization by AO}
\label{alg:pwrmin}
\begin{algorithmic}[1]
\STATE 
$\ell \leftarrow 0$, initialize $\B{p}^{(0)}_i,\,\forall i$
\REPEAT 
\STATE $\ell \leftarrow \ell+1$,
\text{execute commands for all}\ $k \in\{1,\dots,K\}$
\FOR {$m=1$ to $M$}
\STATE \mbox{}\!\!$f_k^{(\ell,m)}\!\leftarrow\!
f_k^{\MMSE,(\ell,m)}$ \hfill [see (\ref{eq:Fmmse})]
\ENDFOR 
\STATE $t_k^{(\ell,m)} \leftarrow$ BC-to-MAC conversion
\hfill [see Sec.~\ref{sec:duality}]
\STATE $\xi_k^{(\ell-1)}\leftarrow \frac{1}{M}\sum_{m=1}^{M}|t_k^{(\ell,m)}|^2$
\STATE  $\bs{T}_k^{(\ell)}\leftarrow \frac{1}{\sqrt{\xi_k^{(\ell-1)}}} \diag(t_k^{(\ell,1)},\ldots,t_k^{(\ell,M)})$
\STATE $\xi_k^{(\ell)}\leftarrow 2^{\rho_k}\,I_k(\bs{\xi}^{(\ell-1)})$ \hfill [power update]
\STATE $\tilde{\B{g}}_{k}^{(\ell)} \leftarrow$ update MAC receiver \hfill [see \eqref{eq:gmmse}]

\FOR {$m=1$ to $M$}
\STATE $t_k^{(\ell,m)} \leftarrow
	\sqrt{\xi_k^{(\ell)}}[\bs{T}_k^{(\ell)}]_{m,m}$ \hfill [include power allocation]

\ENDFOR 	
\STATE $\B{p}_{k}^{(\ell)}\leftarrow$ MAC to BC conversion
\hfill [see Sec.~\ref{sec:duality}]
\UNTIL{$||\bs{\xi}^{(\ell)}-\bs{\xi}^{(\ell-1)}||_1\leq \delta$} 
\end{algorithmic}
\end{algorithm}

\section{Problem Feasibility} 
 \label{sec:feasibility}
\newcommand{\mytkH}{\B{t}_k\:\!\!(\B{H}_k)}
\newcommand{\mytkHone}{\B{t}_k\:\!\!(\B{H}_k^{(1)})}
\newcommand{\mytkHH}{\B{t}_k^{\He}\:\!\!(\B{H}_k)}
\newcommand{\mytiH}{\B{t}_i\:\!\!(\B{H}_i)}
\newcommand{\mytiHH}{\B{t}_i^{\He}\:\!\!(\B{H}_i)}

In this Section we analyze the feasibility of the power minimization problem \eqref{eq:BCreform}. Due to the imperfect CSI assumption, interferences cannot be completely removed in the BC. Consequently, increasing the total transmit power does not necessarily lead to a reduction of the MMSEs for all the users because, although it increases the received power, it also increases the power of the interferences. In certain scenarios, the QoS constraints may require that some users achieve low MMSE values that may be unfeasible even though the transmit power is increased unlimitedly. In the following we present a feasibility test to determine whether it is possible or not to accomplish the QoS constraints $\overline{\MMSE}_k^{\text{MAC}}=2^{-\rho_k}$.

Let us start considering the average MMSE in the MAC
\begin{equation}
\overline{\MMSE}_k^{\text{MAC}}=1-\bar{\bs{\theta}}_k^\He\left(\sigma^2\B{I}_N+\sum_{i=1}^K\Exp[\left|t_i\right|^2\bs{\theta}_i\bs{\theta}_i^\He|\,v]\right)^{-1}\bar{\bs{\theta}}_k,
\label{eq:MMSEMACred}
\end{equation} 
where $\bar{\bs{\theta}}_k=\Exp[\bs{\theta}_kt_k|\,v]$ and $\sigma^2$ is the thermal noise variance in the dual MAC. We now introduce the matrix $\B{\Upsilon}=[\bs{\theta}_1,\dots,
\bs{\theta}_K]\diag(t_1,\ldots, t_K)$ and rewrite \eqref{eq:MMSEMACred} as follows
\begin{align}
\overline{\MMSE}_k^{\text{MAC}}&=1-\left[
\Exp[\B{\Upsilon}^\He|v]\right.\\
&\left.\left(\Exp[\B{\Upsilon}\B{\Upsilon}^\He|v]
+\sigma^2\mathbf{I}_N\right)^{-1}\Exp[\B{\Upsilon}|v]\right]_{k,k}.\nonumber
\end{align}
Hence, the sum average MMSE is
\begin{align}
&\sum_{i=1}^{K}\overline{\MMSE}_i^{\text{MAC}}=K - \label{eq:sumMMSE}\\
&\trace\left(
	\Exp[\B{\Upsilon}^\He|v]\left(\Exp[\B{\Upsilon}\B{\Upsilon}^\He|v]
	+\sigma^2\mathbf{I}_N\right)^{-1}\Exp[\B{\Upsilon}|v]\right).\nonumber	
\end{align}
When $K\geq N$ and the channel knowledge is perfect at both sides, \eqref{eq:sumMMSE} can be made arbitrarily small \cite{HuJo10}. However, due to the imperfect CSI at the MAC receiver we cannot reduce the average MMSE as much as desired. 

Expression \eqref{eq:sumMMSE} allows to determine the region where the feasible average MMSEs lie. Indeed, setting the MAC thermal noise variance to zero (i.e., $\sigma^2=0$) we obtain the following lower bound for the sum average MMSE for any finite total average power allocation
\begin{equation}
	\sum_{i=1}^{K}\overline{\MMSE}^{\text{MAC}}_i>K-\trace\{\B{X}\},
	\label{equation:summsebound}
\end{equation}
where $\B{X}=\Exp[\B{\Upsilon}^\He|v](\Exp[\B{\Upsilon}\B{\Upsilon}^\He|v])^{-1}\Exp[\B{\Upsilon}|v]$. The bound is asymptotically achieved when the powers for all users reach infinity. Therefore, we can formulate a necessary condition for the feasibility of QoS targets: any power allocation with finite sum power achieves an MMSE tuple $\{\overline{\MMSE}_i^{\text{MAC}}\}_{i=1}^K$ inside the polytope 
\begin{equation}
	\mathcal{P}=\left\{\{\overline{\MMSE}_i^{\text{MAC}}\}_{i=1}^K\,|\,\sum_{i=1}^{K}\overline{\MMSE}^{\text{MAC}}_i
	\geq K-\trace\left\{\B{X}\right\}\right\}.
  \label{equation:polytope}
\end{equation}

We now show that for each MMSE tuple in $\mathcal{P}$ there exists a power allocation vector $\bs{\xi}$. To do so, we leverage on the uniqueness property of the fixed point in the interference functions, meaning that if the fixed point exists it is unique and, as a consequence, there is a bijective mapping between the power allocation $\bs{\xi}$ and the average MMSE targets.

Let $\B{f}(\B{x};\B{c})$ be a multivariate function that depends on a vector of independent variables $\B{x}$ and a vector of parameters $\B{c}$. Such function has a fixed point $\B{x}=\B{f}(\B{x};\B{c})$ if it satisfies the following set of sufficient conditions \cite{Kennan} 
\begin{align}
&\B{f}(\mathbf{0};\B{c})\geq\mathbf{0}, 
\label{eq:1req}\\
&\exists\,\B{a}>\mathbf{0}\quad\text{such that}\quad\B{f}(\B{a};\B{c})>\B{a},
\label{eq:2req}\\
&\exists\,\B{b}>\B{a}\quad\text{such that}\quad\B{f}(\B{b};\B{c})<\B{b}.
\label{eq:3req}
\end{align}

We now define $\varepsilon_k=2^{-\rho_k}$ as the MMSE targets in the MMSE QoS constraints \eqref{eq:MMSEconstr} and $\bs{\varepsilon}=[\varepsilon_1,\ldots,\varepsilon_K]^{\Tr}$ as the vector that collects all such targets. We also introduce the following definitions
\begin{align}
\bs{\varphi}_k & =\frac{1}{\sqrt{\xi_k}}\bar{\bs{\theta}}_k, \\ 
\B{\Phi}_k & = \frac{1}{\xi_k} \Exp[(\bs{\theta}_kt_k-\bar{\bs{\theta}}_k)(\bs{\theta}_k t_k-\bar{\bs{\theta}}_k)^\He|\,v], \\ \B{A}_k & =\sum_{i=1}^K\xi_i\B{\Phi}_i+\sum_{j\neq k}\xi_j\bs{\varphi}_j\bs{\varphi}_j^{\He}+\sigma^2\mathbf{I}_N,
\label{A_k}
\end{align}
which, applying the matrix inversion lemma, enable us to rewrite \eqref{eq:MMSEMACred} as
\begin{equation}
	\overline{\MMSE}_k^{\text{MAC}}=\left(1+\xi_k\bs{\varphi}_k^{\He}\B{A}_k^{-1}\bs{\varphi}_k\right)^{-1},
	\label{equation:rewrittenmse}
\end{equation}
and hence define the following functions
\begin{equation}
	f_k(\bs{\xi};\bs{\varepsilon}):=
	\left(\varepsilon_k^{-1}-1\right)\left(\bs{\varphi}_k^{\He}\B{A}_k^{-1}\bs{\varphi}_k\right)^{-1} \forall k.
	\label{eq:fdef}
\end{equation}

We next show that the fixed points $\xi_k = f_k(\bs{\xi};\bs{\varepsilon})$ correspond to the optimal power allocation vectors $\bs{\xi}^{\text{opt}}$ for which $\overline{\MMSE}_k^{\text{MAC}}=\varepsilon_k$, $\forall k$. To do so, we show in the following that the function $\B{f}(\bs{\xi};\bs{\varepsilon}) = [f_1(\bs{\xi};\bs{\varepsilon}), \ldots, f_K(\bs{\xi};\bs{\varepsilon})]^{\Tr}$ satisfies the fixed point conditions \eqref{eq:1req}, \eqref{eq:2req}, \eqref{eq:3req}.

The first requirement \eqref{eq:1req} is easy to show because when the transmit power is $\bs{\xi} = \mathbf{0}$, the inter-user interference drops out and 
\begin{equation}
	f_k(\mathbf{0};\bs{\varepsilon})
	=\frac{1-\varepsilon_k}{\varepsilon_k}\,
	\frac{\sigma^2}{\left\|\bs{\varphi}_k\right\|_2^2}.
\label{fkfunction}
\end{equation}
Note that $f_k(\mathbf{0};\bs{\varepsilon}) \geq 0$ as long as $0 < \varepsilon_k \leq 1$. Moreover, \eqref{fkfunction} also provides a lower bound for $f_k(\bs{\xi};\bs{\varepsilon})$, i.e., for any $\bs{\xi}\geq\mathbf{0}$
\begin{equation}
f_k(\bs{\xi};\bs{\varepsilon})\geq\frac{1-\varepsilon_k}{\varepsilon_k}\,\frac{\sigma^2}{\|\bs{\varphi}_k\|_2^2}.
\label{fklowerbound}
\end{equation}

The second condition \eqref{eq:2req} is also easy to show. Indeed, let $\bar{a}$ be the minimum element of $\B{f}(\mathbf{0};\bs{\varepsilon})$. Hence, $\B{f}(\bs{\xi};\bs{\varepsilon}) \geq \bar{a}\mathbf{1}$ for any $\bs{\xi}\geq\mathbf{0}$. Note from \eqref{fklowerbound} that $\bar{a}>0$ as long as $\varepsilon_k < 1$. Observe now that the power allocation $\bs{\xi} = a\mathbf{1}$ with $a<\bar{a}$ gives $\B{f}(a\mathbf{1};\bs{\varepsilon})\geq\bar{a}\mathbf{1}>a\mathbf{1}$ thus satisfying \eqref{eq:2req}. 

The proof for the last condition \eqref{eq:3req} is more involved and can be found in Appendix \ref{sec:3rdCondition}. 

In summary,	the power minimization problem \eqref{eq:BCreform} has a solution, i.e., the MMSE QoS targets $\bs{\varepsilon}=[2^{-\rho_1},\dots,2^{-\rho_K}]^\Tr$ are feasible, if and only if $\bs{\varepsilon}\in\mathcal{P}$, with $\mathcal{P}$ defined in \eqref{equation:polytope}.

\section{Rate Balancing}
\label{sec:balancing}

So far we have considered the design of the precoders and receivers in a MISO BC to minimize the transmit power while fulfilling certain QoS constraints. However, when the QoS constraints are rather stringent, the problem may be unfeasible. We now address a different problem referred to as rate balancing in the literature, in which the per-user average rate constraints $\{\rho_k\}_{k=1}^K$ are scaled by a common factor $\varsigma \in \mathbb{R}^+$, and a power restriction $P_{\text{tx}}$ is imposed. Observe that, unlike the power minimization formulation, we can relax the per-user requirement so that the problem is always feasible. For such a formulation, we propose to jointly optimize the balance level $\varsigma$ together with the precoders and receivers for given transmit power $P_{\text{tx}}$. 

Using the lower bound \eqref{eq:MMSElb}, the rate balancing problem formulation reads as
\begin{align}
\max_{\{\varsigma(v),\B{p}_k(v),f_k(\B{h}_k)\}_{k=1}^K} &\Exp\left[\varsigma(v)\right]\,\,\text{s.t.}\,\, \Exp\left[\sum_{i=1}^{K}\left\|\B{p}_i(v)\right\|^2_2\right]\leq P_{\text{tx}},\nonumber\\
\text{and}\quad&\hspace{-0.1cm}\Exp\left[\MSE_k^{\text{BC}}\right]\leq 2^{-\Exp\left[\varsigma(v)\right]\rho_k},\,\forall k.
\label{eq:mmseBal}
\end{align}
Following an argumentation similar to the one presented in Sections \ref{sec:sysmodel} and \ref{sec:probForm}, the problem \eqref{eq:mmseBal} can be solved pointwise for each $v$ using the MSE duality and the interference functions. Hence, we rewrite \eqref{eq:mmseBal} as 
\begin{equation}
\max_{\{\varsigma,\xi_k,\B{g}_k\}_{k=1}^K} \varsigma\quad\text{s.t.}\quad \frac{I_k\left(\bs{\xi}\right)}{\xi_k} \leq 2^{-\varsigma \rho_k},\,\text{and}\,\sum_{i=1}^{K}\xi_i\leq P_{\text{tx}}.
\label{eq:balancing}
\end{equation}
where $\bs{\xi} = [\xi_1, \ldots, \xi_K]^T$ is the power allocation vector, $\B{g}_k$ are the dual MAC receivers and $I_k\left(\bs{\xi}\right)$ are the interference functions as given by \eqref{eq:interf}. Similarly to \eqref{eq:macform}, this formulation considers given MAC precoders. Algorithm \ref{alg:pwrmin} can be used to determine optimum filters for given $\varsigma$ but it does not provide the optimum $\varsigma$. Our proposal is to combine it with a bisection search to solve \eqref{eq:balancing}.

Indeed, let us start setting two feasible rate balancing values $\varsigma^\text{L}$ and $\varsigma^\text{H}$ such that $\varsigma^\text{L} \leq \varsigma^\text{opt} \leq \varsigma^\text{H}$. Let $\bs{\xi}^\text{L}$ and $\bs{\xi}^\text{H}$ be the optimum power allocation vectors corresponding to $\varsigma^\text{L}$ and $\varsigma^\text{H}$, respectively. Such optimal power allocation vectors satisfy on the one hand $\frac{I_k(\bs{\xi}^\text{L})}{\xi_k^\text{L}} = 2^{-\varsigma^\text{L} \rho_k}$ and $\frac{I_k(\bs{\xi}^\text{H})}{\xi_k^\text{H}} = 2^{-\varsigma^\text{H} \rho_k}$, and on the other $\sum_{i=1}^{K}\xi_i^\text{L}\leq \sum_{i=1}^{K}\xi_i^{\text{opt}}\leq\sum_{i=1}^{K}\xi_i^\text{H}$, as we will show in the following. 

Now, we introduce the average MMSE balancing factors $\epsilon_k = \frac{2^{-\varsigma\rho_k}}{2^{-\rho_k}} = 2^{-\rho_k(\varsigma-1)}$. Note that increasing the balance level $\varsigma$, decreases the scaling factors $\epsilon_k,\,\forall k$. Let $\epsilon_k^\text{L}$ and $\epsilon_k^\text{H}$ be the MSE scaling factors correspoding to  $\varsigma^\text{L}$ and $\varsigma^\text{H}$, respectively. Note that $\epsilon_k^\text{L}\geq\epsilon_k^\text{opt}\geq\epsilon_k^\text{H}$.

To proof that a bisection search can be performed, we consider $\epsilon_k^\text{L}=a\epsilon_k^{\text{opt}}$, with $a>1$. The constraints in \eqref{eq:balancing} are fulfilled with equality when $\epsilon_k = \epsilon_k^{\text{opt}}$ and $\bs{\xi} = \bs{\xi}^\text{opt}$. Hence, $a\epsilon_k^{\text{opt}}2^{-\rho_k}=a \frac{I_k(\bs{\xi}^\text{opt})}{\xi_k^{\text{opt}}}$ meaning that increasing the MSE targets results in a decrease in the transmit power (i.e. $\xi_k = a^{-1} \xi_k^{\text{opt}}$, $\forall k$) when we keep the interference constant. Moreover, notice that keeping the interference constant sets an upper bound for the interference with the reduced transmit powers $I_k(a^{-1}\bs{\xi}^\text{opt})< I_k(\bs{\xi}^\text{opt})$. Therefore, the power needed to fulfill the constraint with equality is lower than  $a^{-1} \bs{\xi}^\text{opt}$, and $\mathbf{1}^\Tr\bs{\xi}^\text{L}<a^{-1} \mathbf{1}^\Tr\bs{\xi}^\text{opt}< P_{\text{tx}}$ holds. 

We now prove the relationship in the reverse direction, that is, a power reduction translates into larger scaling factors $\epsilon_k$. Let us consider the power reduction $\B{A}\bs{\xi}^\text{opt}$ with $\B{A}=\diag(a_1,\cdots,a_K)<\mathbf{I}$, that leads to certain average MSE scaling factor $\tilde{\epsilon}_k$ for some user $k$, i.e.,  $\tilde{\epsilon}_k2^{-\rho_k}=\frac{1}{a_k\xi_k^\text{opt}} I_k(\B{A}\bs{\xi}^\text{opt})$. Since no assumption about user $k$ has been made, we can focus on user $k^\prime$ such that $a_{k^\prime}\leq a_k\,\forall k$. Consequently, 
\begin{align}
\tilde{\epsilon}_{k^\prime} 2^{-\rho_{k^\prime}}& = \frac{I_{k^\prime}\left(\B{A}\bs{\xi}^\text{opt}\right)}{a_{k^\prime}\xi_{k^\prime}^\text{opt}} \geq \frac{I_{k^\prime}\left(a_{k^\prime}\bs{\xi}^\text{opt}\right)}{a_{k^\prime}\xi_{k^\prime}^\text{opt}} >\nonumber\\
&\frac{I_{k^\prime}\left(\bs{\xi}^\text{opt}\right)}{\xi_{k^\prime}^\text{opt}} = \epsilon_{k^\prime}^\text{opt}2^{-\rho_{k^\prime}}.
\end{align}
Therefore, $\tilde{\epsilon}_{k^\prime} > \epsilon_{k^\prime}^\text{opt}$ for $\bs{\xi}^\text{opt}>\B{A}\bs{\xi}^\text{opt}$. We have previously shown that relaxing the balancing level $\epsilon^\text{opt}_k$ implies a power reduction with respect to $\bs{\xi}^\text{opt}$. Hence, we conclude that a power reduction entails a lower balancing level $\varsigma$, and vice-versa, when the precoders, receive filters, and power allocation vectors are optimum for every balancing level.

Finally, reducing the gap between $\varsigma^\text{L}$ and $\varsigma^\text{H}$ results in the optimum balancing level $\varsigma^\text{opt}$ for the total average transmit power $\mathbf{1}^\Tr\bs{\xi}^\text{opt}=P_{\text{tx}}$.

\subsection{Rate Balancing Algorithm}
Algorithm \ref{alg:balancing} presents the steps to solve the optimization problem \eqref{eq:balancing}. The algorithm is initialized with two balancing levels $\varsigma^{\text{L},(0)}$ and $\varsigma^{\text{H},(0)}$ (line 1). Next, their corresponding vector power allocation vectors, $\bs{\xi}^{\text{H},(0)}$ and $\bs{\xi}^{\text{L},(0)}$, are computed via Algorithm \ref{alg:pwrmin} (line 2). Observe that the optimum lies in between the initial balance levels. Next, the algorithm enters a loop that first computes a new balancing level as the geometric mean of the balancing levels obtained in the previous iteration (line 5). Then, the power allocation vector for this new balancing level is computed via Algorithm \ref{alg:pwrmin} (line 6). Next, we check whether the power obtained is lower than the power constraint or not (line 7) and update the balancing levels accordingly (lines 8 and 10). Finally, we test if the current power has the desired accuracy (line 12).

The proof for the convergence of Algorithm \ref{alg:balancing} depends on the feasibility of the initial average MSE targets $2^{-\varsigma^{\text{H},(0)}\rho_k}\,\forall k$. Indeed, recall that the feasibility region is described in Section \ref{sec:feasibility} as a bounded polytope and that the initial balancing levels $\varsigma^{\text{L},(0)}$ and $\varsigma^{\text{H},(0)}$ are chosen such as $\varsigma^{\text{L},(0)}\leq\varsigma^\text{opt}\leq\varsigma^{\text{H},(0)}$. Hence, if $2^{-\varsigma^{\text{H},(0)}\rho_k}\,\forall k$ lies inside the polytope so does $2^{-a\varsigma^{\text{H},(0)}\rho_k}\,\forall k$ for any $0\leq a<1$. Taking into account that the average MMSE given by $\frac{1}{\xi_k^{(\ell)}}I_k(\bs{\xi}^{(\ell)})$ is monotonically decreasing in $\bs{\xi}^{(\ell)}$, the bisection procedure reduces the gaps $(\varsigma^{\text{H},(\ell)}-\varsigma^{\text{L},(\ell)})$ and $|\mathbf{1}^\Tr\bs{\xi}^{(\ell)}-P_{\text{tx}}|$ at every iteration until a desired accuracy $\delta$ is achieved.

\begin{algorithm}[t]
\caption{Rate Balancing}
\label{alg:balancing}
\begin{algorithmic}[1]
\STATE $\ell \leftarrow 0$, initialize $\varsigma^{\text{L},(0)},\varsigma^{\text{H},(0)}$
\STATE find $\bs{\xi}^{\text{H},(0)}\leq \bs{\xi}^{\text{L},(0)}$ via Alg. \ref{alg:pwrmin} \hfill [power min.]
\REPEAT
\STATE $\ell\leftarrow\ell+1$
\STATE $\varsigma^{(\ell)}\leftarrow\sqrt{\varsigma^{\text{L},(\ell-1)}\varsigma^{\text{H},(\ell-1)}}$ \hfill [new candidate]
\STATE find $\bs{\xi}^{(\ell)}$ for $\varsigma^{(\ell)}$ via Alg. \ref{alg:pwrmin} \hfill [power min.]
\IF {$\sum_{i=1}^{K}\xi^{(\ell)}_i<P_{\text{tx}}$}
\STATE $\varsigma^{\text{H},(\ell)}\leftarrow\varsigma^{(\ell)}$, $\varsigma^{\text{L},(\ell)}\leftarrow\varsigma^{\text{L},(\ell-1)}$ \hfill[weights update]
\ELSE
\STATE $\varsigma^{\text{L},(\ell)}\leftarrow\varsigma^{(\ell)}$, $\varsigma^{\text{H},(\ell)}\leftarrow\varsigma^{\text{H},(\ell-1)}$
\hfill[weights update]
\ENDIF
\UNTIL{$|\sum_{i=1}^{K}\xi^{(\ell)}_i-P_{\text{tx}}|<\delta$}
\end{algorithmic}
\end{algorithm}

\section{Simulation Results}
\label{sec:simul}
In this section we present the results of several simulation experiments carried out to show the performance of the proposed algorithms. First, let us introduce the following error model corresponding to the imperfect CSIT
 \begin{equation}
\B{h}_k=\bar{\B{h}}_k+\tilde{\B{h}}_k,
\label{eq:channel_model}
\end{equation}
where $\bar{\B{h}}_k=\Exp[\B{h}_k|\,v]$ and $\tilde{\B{h}}_k$ is the error. This flexible model can represent, for example, the errors due to calibration in TDD systems or the quantization and estimation errors in FDD systems. We assume that the imperfect CSI error is zero-mean Gaussian, i.e. $\tilde{\B{h}}_k\sim\mathcal{N}_\mathbb{C}(\mathbf{0},\B{C}_k)$ where $\B{C}_k=\Exp[(\B{h}_k-\bar{\B{h}}_k)(\B{h}_k-\bar{\B{h}}_k)^\He|\,v]$ is the $k$-th user CSI error covariance matrix. Recall that $v$ and $f_{\B{h}_k|v}(\B{h}_k|v)$ are known at the transmitter, although the specific realizations of $\B{h}_k$ and $\tilde{\B{h}}_k$ are not.  According to that assumption, it is possible to generate the channel realizations  $\B{h}_k^{(m)}=\bar{\B{h}}_k+\tilde{\B{h}}_k^{(m)}$ for $k=\{1,\ldots,K\}$ and $m=\{1,\ldots,M\}$, with $\bar{\B{h}}_k=\Exp[\B{h}_k|v]$ and $\tilde{\B{h}}_k^{(m)}\sim\mathcal{N}_{\mathbb{C}}(\mathbf{0},\B{C}_k)$. In our scenario, the number of users and transmit antennas were $K=4$ and $N=4$, respectively. We generated $M=1000$ channel realizations considering $\B{C}_k=\mathbf{I}_N$, and $\bar{\B{h}}_k\sim\mathcal{N}_{\mathbb{C}}(\mathbf{0},\mathbf{I}_N),\forall k$. We also considered $\sigma^2_{\bs{\eta}_k}=1,\,\forall k$.

\subsection{Power Minimization}
\label{sec:PMinRes}
In this subsection the Algorithm \ref{alg:pwrmin} that solves the optimization problem \eqref{eq:powermin} is considered. We choose users with different rate requirements, viz., $\rho_1=0.5146$, $\rho_2=0.737$, $\rho_3=1$ and $\rho_4=0.2345$ bits per channel use, respectively. These requirements correspond to the following targets in the MMSE domain: $\varepsilon_1 = 0.7$, $\varepsilon_2=0.6$, $\varepsilon_3=0.5$ and $\varepsilon_4=0.85$. The threshold in Algorithm \ref{alg:pwrmin} is set to $\delta=10^{-2}$. Initial precoders are random. 

\begin{figure}[ht]
\begin{minipage}{1.0\linewidth}
\includegraphics[width=0.95\linewidth]{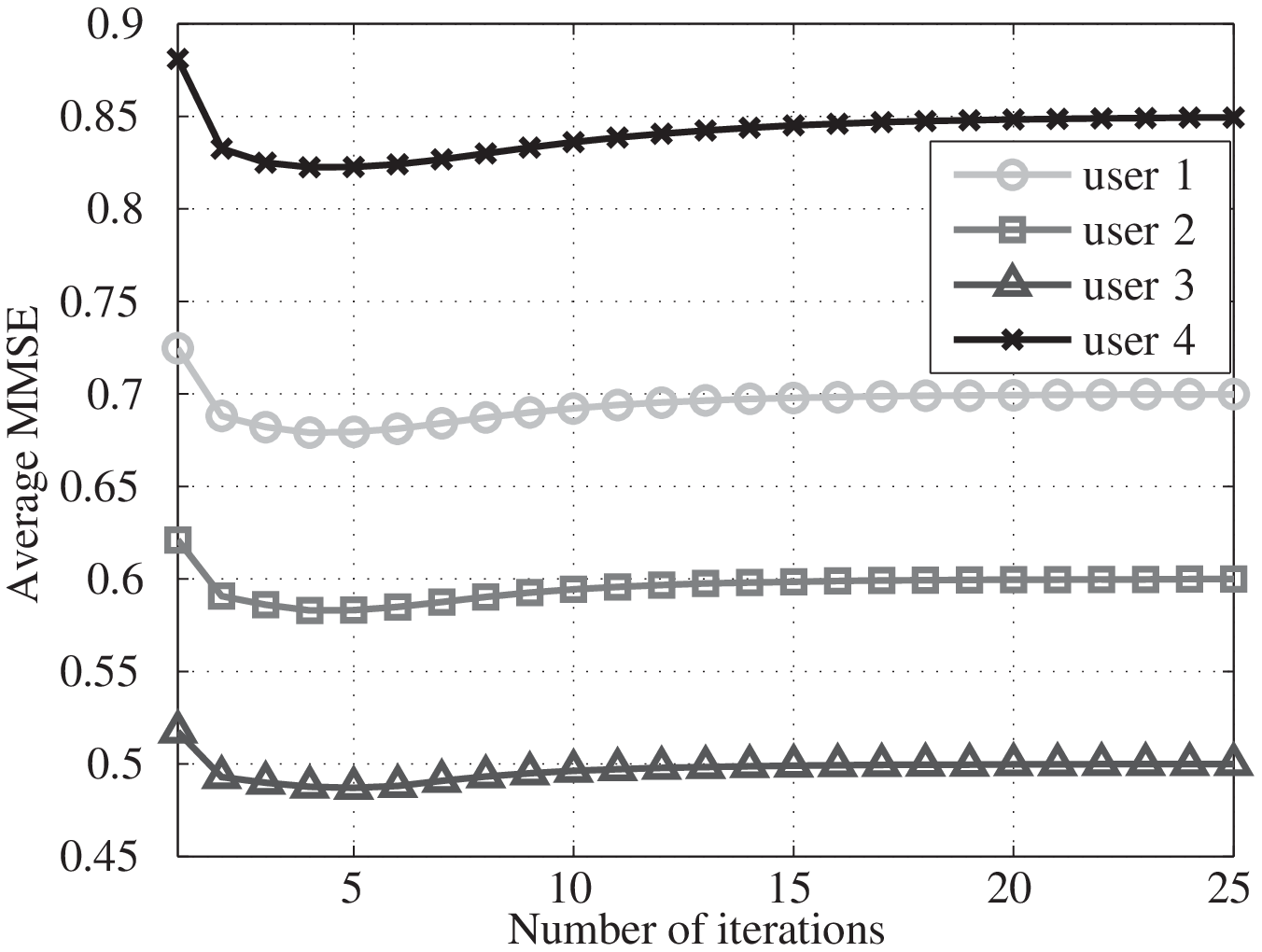}
\caption{Power Minimization: MMSEs vs. Number of Iterations.}
\label{fig:MMSEvs}
\end{minipage}
  
\begin{minipage}{1.0\linewidth}
\includegraphics[width=0.95\linewidth]{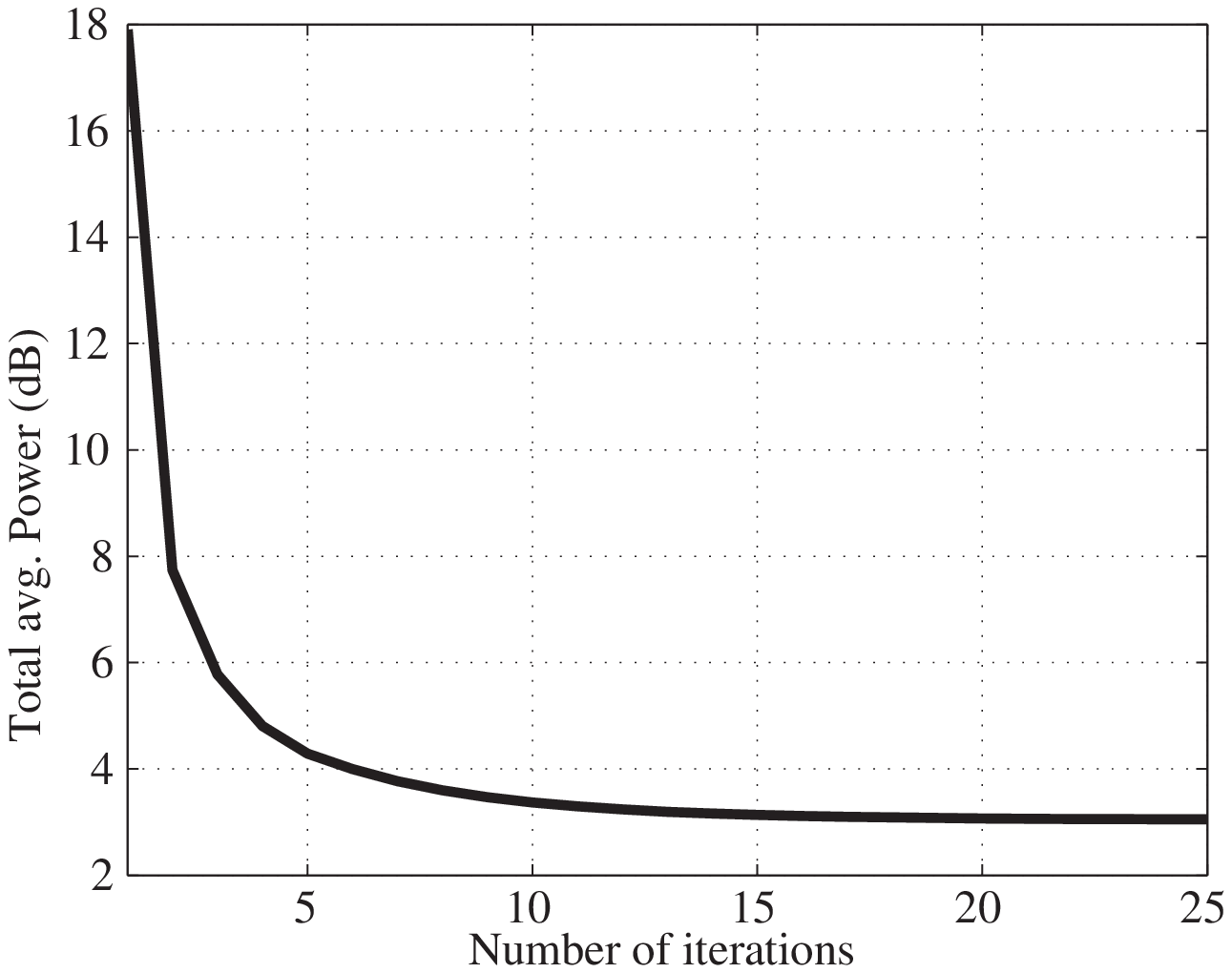}
\caption{Power Minimization: Total Average Power vs. Number of Iterations.}
\label{fig:PwrVs}
\end{minipage}

\begin{minipage}{1.0\linewidth}
\includegraphics[width=0.95\linewidth]{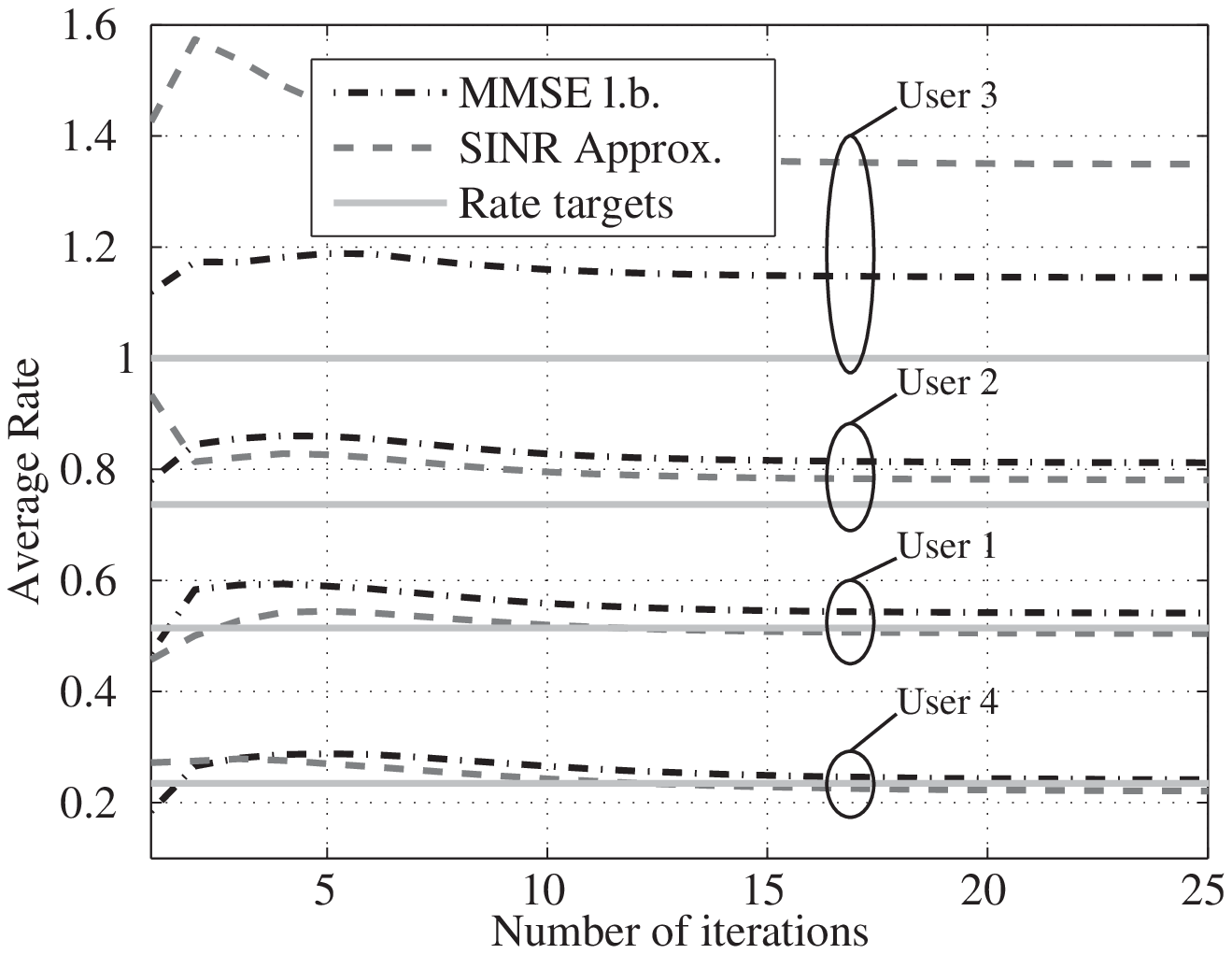}
\caption{Power Minimization: Rates vs. Number of Iterations.}
\label{fig:rateVs}
\end{minipage}
\end{figure}

Fig. \ref{fig:MMSEvs} shows how the MMSE for all users converges to the desired targets $\varepsilon_k$. Since the problem is feasible, the minimum total average power will be reached when the constraints in \eqref{eq:macform} are fulfilled with equality. As can be seen in Fig. \ref{fig:MMSEvs}, the first steps go in the direction of fulfilling the requirements and the MMSEs are reduced. Nevertheless, the subsequent iterations increase the MMSEs until the targets $\varepsilon_k$ are reached for all users. Correspondingly, as shown in Fig. \ref{fig:PwrVs}, the total average power is initially above $16$ dB and it gradually reduces throughout the subsequent iterations until convergence is reached at $3$ dB. The total average power is dramatically reduced during the first five iterations whereas the improvement is marginal after iteration $15$.

Fig. \ref{fig:rateVs} shows the evolution of the average rates over the iterations. Recall from \eqref{eq:MMSElb} that the actual average rates are lower bounded by the MMSE-based targets $\varepsilon_k$, i.e., $\Exp[R_k|\,v] \geq -\log_2(\varepsilon_k)$, as discussed in Section \ref{sec:sysmodel} [see \cite{ShLa08}]. The gap between the average rates obtained with Algorithm \ref{alg:pwrmin} and the average rate targets corresponding to the QoS constraints can be also observed from Fig. \ref{fig:rateVs}. Moreover, we also include in this figure the rates obtained employing the SINR approximation utilized in \cite{ScBo04} and widely employed afterwards (e.g. \cite{ShLa08},\cite{ChLi07},\cite{KoCa07}). This approach determines the average rates as $\log_2(1+\overline{\text{SINR}}_k)$ where $\overline{\text{SINR}}_k$ is obtained from applying separately the expectation operator to both the numerator and the denominator of the SINR, i.e.,
\begin{equation}
\overline{\text{SINR}}_k=\frac{\B{p}_k^\He\Exp\left[\B{h}_k\B{h}_k^\He|\,v\right]\B{p}_k}{\sigma_{\bs{\eta}_k}^2+\sum_{i\neq k}\B{p}_i^\He\Exp\left[\B{h}_k\B{h}_k^\He|\,v\right]\B{p}_i}.
\end{equation}     
Fig. \ref{fig:rateVs} shows the resulting values for $\log_2(1+\overline{\text{SINR}}_k)$ along the iterations in Algorithm \ref{alg:pwrmin}. Note that the average rates for the SINR approximation are larger than the true average rates for users $2$ and $3$, but smaller for users $1$ and $4$. Hence, it is not possible to guarantee the QoS restrictions. Contrary to this, fulfilling the MMSE-based targets, as proposed in our approach, ensures average rates larger than the targets.  

\subsection{Rate Balancing}
This subsection focuses on the performance of Algorithm \ref{alg:balancing}. This algorithm solves the optimization problem \eqref{eq:mmseBal} by means of Algorithm \ref{alg:pwrmin} and a bisection process for which it is necessary to decide two starting points, $\varsigma^{\text{L},(0)}$ and $\varsigma^{\text{H},(0)}$, such that the optimum balancing level lies in between, i.e., $\varsigma^{\text{L},(0)}\leq\varsigma^{\text{opt}}\leq\varsigma^{H,(0)}$. The rate targets employed in Subsection \ref{sec:PMinRes} are also used in this section. We scale them with a common factor to obtain the rate targets. The threshold to check convergence is set to $\delta=10^{-2}$.

\begin{figure}[ht]
\includegraphics[width=0.95\linewidth]{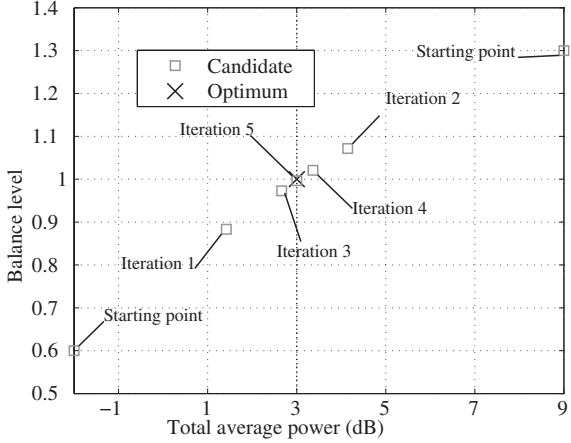}
\caption{MMSE balancing: Balancing Level Candidates vs. Total Average Power in dB.}
\label{fig:balancing}
\end{figure}

Taking into account the numerical results obtained in Subsection \ref{sec:PMinRes}, we consider a total average transmit power of $3$ dB leading to an expected balancing level of approximately one. Therefore, we pick $\varsigma^{\text{L},(0)}=0.6$ and $\varsigma^{\text{H},(0)}=1.3$, from which $\varsigma^{\text{opt}}\in[0.6, 1.3]$. Fig. \ref{fig:balancing} plots the average power versus the balancing level for the different iterations of the bisection algorithm. The two initial values correspond to the points located on the left and the right vertical axis in the figure. Note that the searching interval reduces as the algorithm progresses until it converges after five iterations to the point $\varsigma^\text{opt}=0.99659$ and $P_{\text{tx}}=3.0072$ dB. This is in accordance to the experimental results obtained in Subsection \ref{sec:PMinRes}. 

\begin{figure}[ht]
\begin{minipage}{1.0\linewidth}
\includegraphics[width=0.95\linewidth]{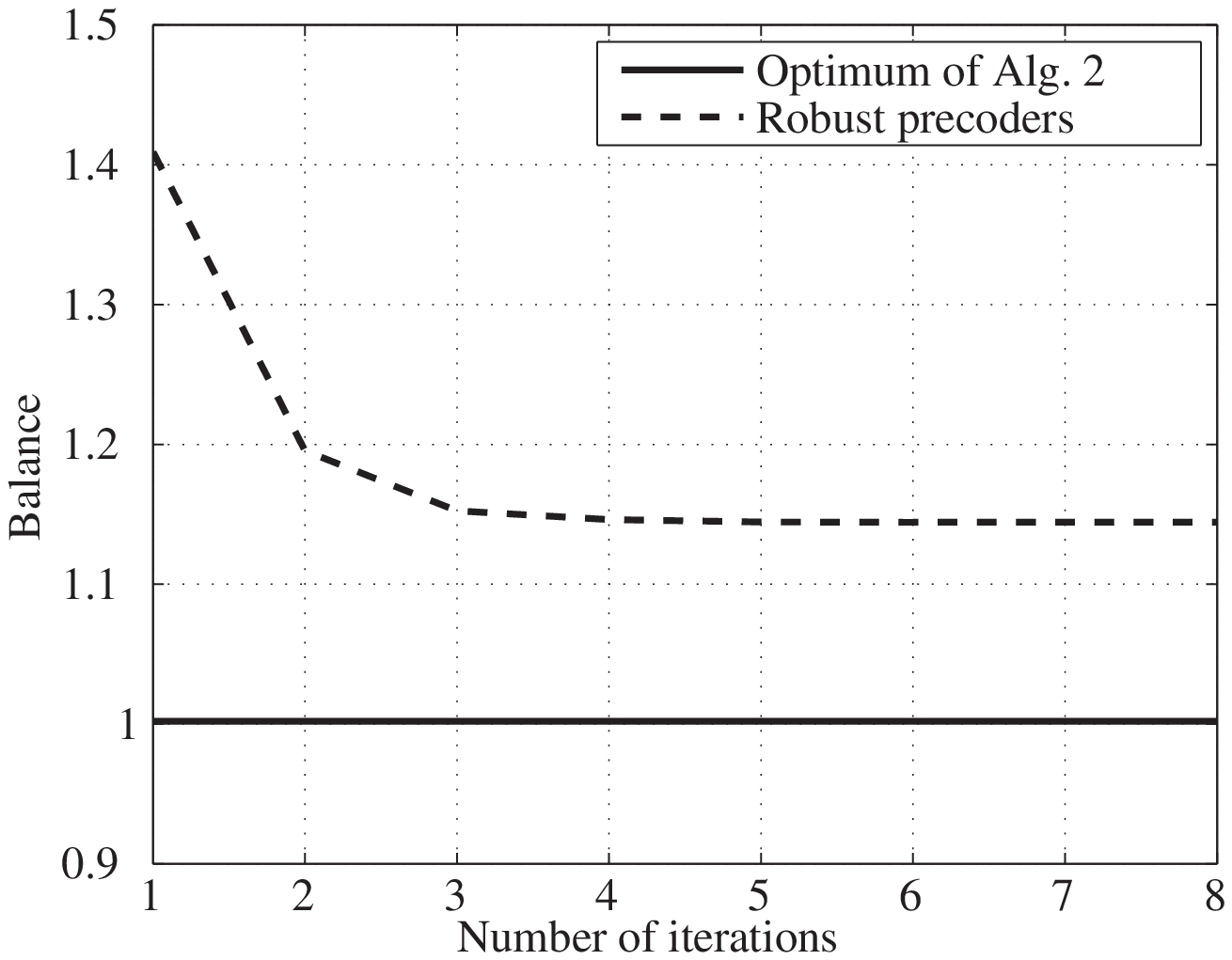}
\caption{Robust Transceiver: Balancing Level vs. Number of Iterations.}
\label{fig:RobTranscConverg}
\end{minipage}
  
\begin{minipage}{1.0\linewidth}
\includegraphics[width=0.95\linewidth]{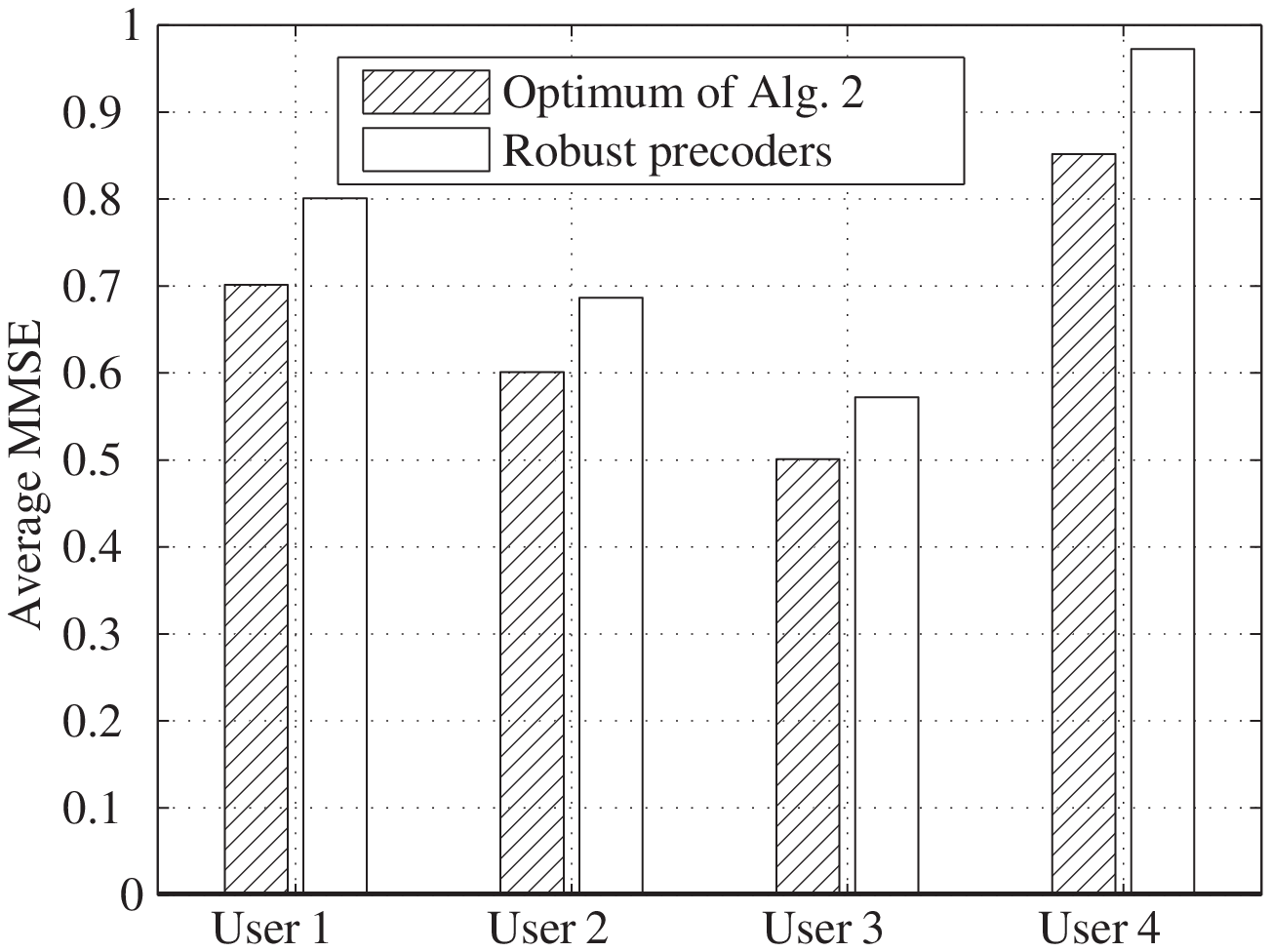}
\caption{Robust Transceiver: Average MMSEs for proposed Alg. \ref{alg:balancing} vs. Average MMSEs for Robust Transceivers.}
\label{fig:RobTrVS}
\end{minipage}

\end{figure}

We also performed a computer experiment to compare our approach to that presented in \cite{BoChVa11}, where a duality that allows to solve several optimization problems considering a scenario where the users and the BS share the same CSI. More specifically (see Section V.B of \cite{BoChVa11}) the following weighted MSE Min-Max problem is addressed
\begin{equation}
\min_{\{\B{p}_k,\B{f}_k\}_{k=1}^K}\max_i \frac{\overline{\MSE}_i^{\text{BC}}}{w_i}\quad\text{s.t.}\quad\sum_{j=1}^{K}\left\|\B{p}_j\right\|^2_2\leq P_{\text{tx}}
\label{eq:MinMax}
\end{equation}
where $w_i$ is the weight for the $i$th user. Robust precoders and filters are designed via an AO process, and the power allocation is calculated solving an eigen-system \cite{ShScBo08}. The optimum of \eqref{eq:MinMax} is obtained after a few iterations and fulfills  $\sum_{i=1}^{K}\left\|\B{p}_i\right\|^2_2=P_{\text{tx}}$ and  $\overline{\MMSE}_k^{\text{BC}}/w_k=w^\text{opt},\,\forall k$ (see Fig. \ref{fig:RobTranscConverg}). The error precision for the min max ratio $w^\text{opt}$ is $10^{-4}$.

This min max problem can be seen as a balancing problem with $w_i=\varepsilon_i$. Thus, Fig. \ref{fig:RobTrVS} represents the comparison between the solutions employing robust transceivers and the one proposed in this work. As can be seen in the figure, the proposed Algorithm \ref{alg:balancing} performs better because $\varsigma^\text{opt}=0.99659$ is closer than $w^\text{opt}=1.1442$ to $1$. However, the robust filters from \cite{BoChVa11} are designed sharing imperfect CSI using a computationally cheaper algorithm.

\section{Conclusion}
\label{conclusions}
We focused on the design of linear precoders and receivers to minimize the transmit power in a MISO BC while fullfiling a set of per-user QoS constraints expressed in terms of per-user average rate requirements. We explained that QoS constraints can be substituted by more manageable restrictions based on the average MMSE. We next exploited the MSE BC/MAC duality to jointly determine the optimum transmit and receive filters by means of an Alternating Optimization (AO) algorithm. Additionally, the optimum power allocation is found employing the so-called standard interference functions framework. We also analyzed the problem feasibility to ensure convergence of the proposed algorithm. Moreover, we addressed the balancing problem combining the proposed algorithm with a search. We carried out simulation experiments to show the performance of the proposed methods and compare them with existing solutions in the literature.



\appendices

\section{Average-MMSE-Based Lower Bound Gap}
\label{sec:gap}

In this appendix, we study the gap between the average rate and the average MMSE lower bound in the inequality \eqref{eq:MMSElb}. For simplicity reasons, we focus on the case where $K=N=1$. In such case, the MISO BC  \eqref{hatsk} reduces to a \emph{Single-Input Single-Output} (SISO) system model. Considering $h \sim \mathcal{N}_\mathbb{C}(0,1)$ and $\eta \sim \mathcal{N}_\mathbb{C}(0,\sigma^2)$, the  average MMSE is
\begin{equation}
\Exp[\MMSE] = \Exp \left[\frac{\sigma^2}{|hp|^2+\sigma^2}\right].
\label{eq:expecMMSE}
\end{equation}

\begin{figure}[ht]
\includegraphics[width=\linewidth]{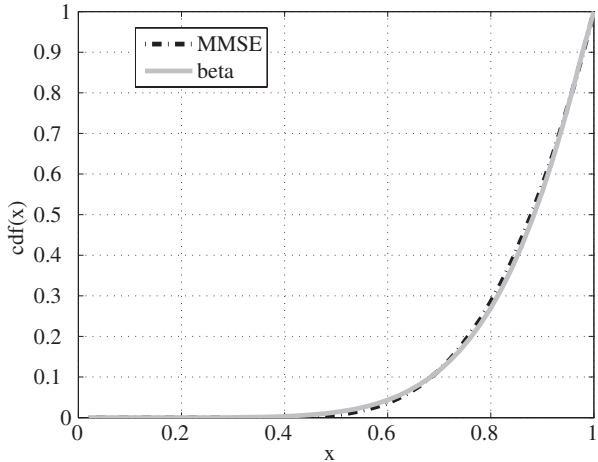}
\caption{MMSE Cumulative Distribution vs. Beta Cumulative Distribution.}
\label{fig:approx}
\end{figure}

We now approximate the MMSE \emph{Cumulative Distribution Function} (CDF) by a beta distribution. Fig. \ref{fig:approx} illustrates the tightness of such approximation showing the CDF of the MMSE for $|p|^2=1$ and $\sigma^2=10$, and the CDF of a beta random variable with $\alpha=6.54162$ and $\beta=1.12133$. 

We next introduce the PDF $f_{\varepsilon}(\MMSE)$ and the auxiliary variable $\varepsilon=\MMSE$. Now, the expectation of the logarithm of $\varepsilon$ is
\begin{equation*}
\Exp[\ln(\varepsilon)]=\int_{0}^{1}f_{\varepsilon}(\varepsilon)\ln(\varepsilon)dx. 
\end{equation*}
Considering $\varepsilon$ has a beta PDF, the logarithm of the geometric mean reads as
\begin{align*}
&\Exp[\ln(\varepsilon)]=\psi(\alpha)-\psi\left(\alpha+\beta\right),
\end{align*}
where $\psi(x)$ is the digamma function. Such a function can be approximated as $\psi(x)\approx \ln(x+\frac{1}{2})$ for $x>1$. 
Then, the average MMSE lower bound is approximated as follows
\begin{align*}
&-\Exp[\log_2(\varepsilon)]\approx \frac{1}{\ln(2)}\log_2\left(1+\frac{\beta}{\alpha-\frac{1}{2}}\right)
\end{align*} 

Considering the expectation of the beta distribution is $\Exp[\varepsilon]=\frac{\alpha}{\alpha+\beta}$, the average MMSE lower bound  is $-\log_2(\Exp[\varepsilon])=\frac{1}{\ln(2)}\ln(1+\frac{\beta}{\alpha})$. Hence, the gap between the average rate $\Exp[R]$ and the lower bound is
\begin{align*}
&\Exp\left[R\right]-\left[-\log_2\left(\Exp\left[\MMSE\right]\right)\right]\\
&\approx\frac{1}{\ln(2)}\log_2\left(1+\frac{\beta}{\alpha-\frac{1}{2}}\right)
-\frac{1}{\ln(2)}\ln\left(1+\frac{\beta}{\alpha}\right)\\
&=\frac{1}{\ln(2)}\left[\ln\left(1+\frac{\beta}{\alpha-\frac{1}{2}}\right)-\ln\left(1+\frac{\beta}{\alpha}\right)\right]\\
&=\log_2\left(1+\frac{\frac{\beta}{2}}{\left(\alpha-\frac{1}{2}\right)\left(\alpha+\beta\right)}\right).
\end{align*}
In our example this gives $\log_2(1+0.0121)=0.0174$.

\section{Interference Function Properties}
\label{ap:intfunc}
We show in this appendix that $I_k(\bs{\xi})$ as given by \eqref{eq:interf} satisfies the properties of a standard interference function.

Observe that $y_k-\xi_k|\tilde{\B{g}}_k^\He\bs{\mu}_k|^2$ with $y_k$ from \eqref{eq:MACscalar} is positive and increasing in $\bs{\xi}$. Then, it is straightforward to see that $I_k(\bs{\xi})$ is positive. Moreover, since both terms inside the outer inverse of \eqref{eq:interf} are decreasing in $\bs{\xi}$, the whole expression increases with $\bs{\xi}$ and satisfies monotonicity.

To prove scalability we consider the scalar $a>1$. Hence
\begin{align}
&aI_k(\bs{\xi})=a\left(\frac{1}{\xi_k}+\left|\tilde{\B{g}}_k^\He\bs{\mu}_k\right|^2\left(y_k-\xi_k\left|\tilde{\B{g}}_k^\He\bs{\mu}_k\right|^2\right)^{-1}\right)^{-1}>\nonumber\\
&\left(\frac{1}{a\xi_k}+\frac{\left|\tilde{\B{g}}_k^\He\bs{\mu}_k\right|^2}{a}\left(z_k-\xi_k\left|\tilde{\B{g}}^\He_{k}\bs{\mu}_k\right|^2\right)^{-1}\right)^{-1}=I_k(a\bs{\xi}),\nonumber
\end{align}
where $z_k=\tilde{\B{g}}_k^\He(\sum_{i=1}^{K}\xi_i\B{\Theta}_i+\frac{1}{a}\mathbf{I}_N)\tilde{\B{g}}_k$.
%

\section{Proof for the condition \eqref{eq:3req}}
\label{sec:3rdCondition}
The proof for the condition \eqref{eq:3req} will be divided into two cases depending on the number of users and transmit antennas. 
\subsubsection{$N\geq K$}
This is the case where the number of transmit antennas is greater than or equal to the number of users. We start searching for an upper bound for $f_k(\bs{\xi};\bs{\varepsilon})$, or equivalently, a lower bound for the inverse term in \eqref{eq:fdef}. To do so, we introduce the following matrices
\begin{align}
\B{B}_{\bar{k}} & =[\bs{\varphi}_{i_1},\ldots,\bs{\varphi}_{i_{K-1}}]_{i_j\neq k,\,\forall j}, \\ 
\B{\Xi}_{\bar{k}} & =\diag(\xi_i)_{i\neq k},
\end{align}
which allow us to rewrite the second summand in \eqref{A_k} as $\sum_{i\neq k}\xi_i\bs{\varphi}_i\bs{\varphi}_i^{\He} = \B{B}_{\bar{k}}\B{\Xi}_{\bar{k}}\B{B}_{\bar{k}}^{\He}$. If we also define 
\begin{align}
\B{\Phi} & =\sum_{i=1}^K\xi_i\B{\Phi}_i+\sigma^2\mathbf{I}_N,
\end{align}
we can rewrite the matrix $\B{A}_k$ as
\begin{equation}
\B{A}_k=\B{\Phi}+\B{B}_{\bar{k}}\B{\Xi}_{\bar{k}}\B{B}_{\bar{k}}^\He.
\end{equation}
Applying now the matrix inversion lemma it is possible to write the inverse of $\B{A}_k$ as
\begin{equation*}
	\B{A}_k^{-1} = \B{\Phi}^{-1}
	\left[\mathbf{I}_N-\B{B}_{\bar{k}}\left(\B{\Xi}_{\bar{k}}^{-1}+\B{B}_{\bar{k}}^{\He}\B{\Phi}^{-1}
	\B{B}_{\bar{k}}\right)^{-1}\B{B}_{\bar{k}}^{\He}\B{\Phi}^{-1}\right].
\end{equation*}
Defining $\bs{\psi}_k=\B{\Phi}^{-1/2}\bs{\varphi}_k$ and
$\B{D}_{\bar{k}}=\B{\Phi}^{-1/2}\B{B}_{\bar{k}}$ leads us, eventually, to the lower bound 
\begin{equation}
	\bs{\varphi}_k^{\He}\B{A}_k^{-1}\bs{\varphi}_k\geq
  \bs{\psi}_k^\He\left(\mathbf{I}_N-\B{D}_{\bar{k}}\left(\B{D}_{\bar{k}}^{\He}\B{D}_{\bar{k}}
	\right)^{-1}\B{D}_{\bar{k}}^{\He}\right)\bs{\psi}_k,
\label{eq:fuppb}	
\end{equation}
and the corresponding upper bound
\begin{equation*}
f_k(\bs{\xi};\bs{\varepsilon})\leq\frac{1-\varepsilon_k}{\varepsilon_k}\left(\bs{\psi}_k^\He\left(\mathbf{I}_N-\B{D}_{\bar{k}}\left(\B{D}_{\bar{k}}^{\He}\B{D}_{\bar{k}}
	\right)^{-1}\B{D}_{\bar{k}}^{\He}\right)\bs{\psi}_k\right)^{-1}.
\end{equation*}
Notice that matrix $\B{D}_{\bar{k}}^{\He}\B{D}_{\bar{k}}$ is non-singular when $N\geq K$. Observe that the equality in the last expression holds for $\xi_k\rightarrow\infty$, $\forall k$. Since $\B{f}(\bs{\xi};\bs{\varepsilon}) \geq \bar{a}\mathbf{1} > a\mathbf{1}$ for any $\bs{\xi}\geq\mathbf{0}$ sets a lower bound, we only have to find $\B{b}$ such that
$b_k>(\frac{1}{\varepsilon_k}-1)(\bs{\psi}_k^\He(\mathbf{I}-\B{D}_{\bar{k}}(\B{D}_{\bar{k}}^{\He}\B{D}_{\bar{k}})^{-1}\B{D}_{\bar{k}}^{\He})\bs{\psi}_k)^{-1}$ to complete the proof for the third requirement \eqref{eq:3req} when $N\geq K$ .

\subsubsection{$N<K$} We now focus on the case in which the number of transmit antennas is smaller than the number of users. The power allocation is set to $\B{b}=\alpha\B{b}_0$, where $\B{b}_0$ belongs to the simplex
$\mathcal{S}=\{\B{x}\vert\sum_k x_k=1\ \text{and}\  x_k\geq 0 ~ \forall k\}$.
For $\alpha\rightarrow\infty$ (or $\sigma^2\rightarrow 0$)
and $\B{b}_0>\mathbf{0}$, we can rewrite \eqref{eq:fdef} as
\begin{equation*}
	f_k^{\infty}(\B{b}_0;\bs{\varepsilon}):=
	\frac{\frac{1}{\varepsilon_k-1}}
	{\bs{\varphi}_k^{\He}\Big(\sum\limits_i b_{0,i}\B{\Phi}_i
	+\sum\limits_{j\neq k}b_{0,j}\bs{\varphi}_j\bs{\varphi}_j^{\He}\Big)^{-1}
	\bs{\varphi}_k}.
\end{equation*}
The average MMSE targets collected in $\bs{\varepsilon}$ have to satisfy equality in \eqref{equation:summsebound} for $\alpha\rightarrow\infty$, i.e., a tuple $\bs{\varepsilon}$ that lies in the region that separates feasible from unfeasible targets
$\mathcal{B}=\{\bs{\varepsilon}|\mathbf{1}^\Tr\bs{\varepsilon}=
K-\trace(\B{X}) \}$. Note that $\B{b}_0=\B{f}^{\infty}(\B{b}_0;\bs{\varepsilon})$ is a fixed point of $\B{f}^{\infty}$ but we need to verify the bijective mapping in order to complete the proof, that is, for any average MMSE target tuple
$\B{\varepsilon}\in\mathcal{B}$ there is a unique power allocation $\B{b}=\alpha\B{b}_0$.

First, we define the SINR as $\text{SINR}=1/\overline{\MMSE}^\text{MAC}_k-1$. In the limit case $\alpha\rightarrow\infty$, the expression for the  \emph{Signal to Interference Ratio} (SIR) is
\begin{equation*}
	\text{SIR}_k=b_{0,k}\bs{\varphi}_k^{\He}\bigg(
	\sum_{i=1}^K b_{0,i}\B{\Phi}_i
	+\sum_{j\neq k}b_{0,j}\bs{\varphi}_j\bs{\varphi}_j^{\He}
	\bigg)^{-1}\bs{\varphi}_k,
\end{equation*}
from which we rewrite $\text{SIR}_k=b_{0,k}(\mathcal{Q}_k(\B{b}_0))^{-1}$. Thus, we can use the properties of the function  $\mathcal{Q}_k(\B{b}_0)$ (see \cite{BoSc07}) to guarantee the existence and uniqueness of the optimal power allocation for the balancing problem 
\begin{equation} 
	\max_{r,\B{b}_0}\ r\quad\text{s.t.}\quad
	\frac{b_{0,k}}{\mathcal{Q}_k\left(\B{b}_0\right)}=r\,\text{SIR}_{0,k}
	\quad\forall k\in\{1,\dots, K\}.
  \label{eq:sirbalancing}
\end{equation}
Since we established a relationship between the SIR and the $\overline{\MMSE}^\text{MAC}$ when we let the power grow without restriction (i.e. $\alpha\rightarrow\infty$), we use the bound for $\mathbf{1}\bs{\varepsilon}$ to find the optimal balancing level $r$ for \eqref{eq:sirbalancing} via 
\begin{equation}
	\sum_{i=1}^K\frac{1}{1+r\,\text{SIR}_{0,i}}
	=K-\trace\{\B{X}\}.
	\label{eq:siroptimum}
\end{equation}

The previous equation only has a single solution since the functions $(1+r\,\text{SIR}_{0,k})^{-1}$ are monotonically decreasing with $r>0$, e.g., if we obtain the SIR targets from MMSE targets lying in the region of interest $\mathcal{B}$,  \eqref{eq:siroptimum} is fulfilled with $r=1$.

Thus far we have shown that a unique
power allocation $\B{b}=\alpha\B{b}_0$, with $\B{b}_0\in\mathcal{S}$
and $\alpha\rightarrow\infty$, always exists
for any MMSE tuple in the region that separates feasible and unfeasible targets $\bs{\varepsilon}^\prime\in\mathcal{B}$ such that
$\B{f}(\B{b};\bs{\varepsilon}^\prime)=\B{b}$.
Note that $\B{f}(\B{b};\bs{\varepsilon})$ is decreasing in $\bs{\varepsilon}$ and we can prove that the third requirement \eqref{eq:3req} is also fulfilled
for $N<K$ due to the fact that for any target $\bs{\varepsilon}>\bs{\varepsilon}^\prime$ we have
\begin{equation}
  \B{f}(\B{b};\bs{\varepsilon})<\B{b}.
\end{equation}

\ifCLASSOPTIONcaptionsoff
  \newpage
\fi



\bibliographystyle{IEEEtran}
\bibliography{refs}
%
%
%

%

\begin{IEEEbiography}[{\includegraphics[width=1in,height=1.25in,clip,keepaspectratio]{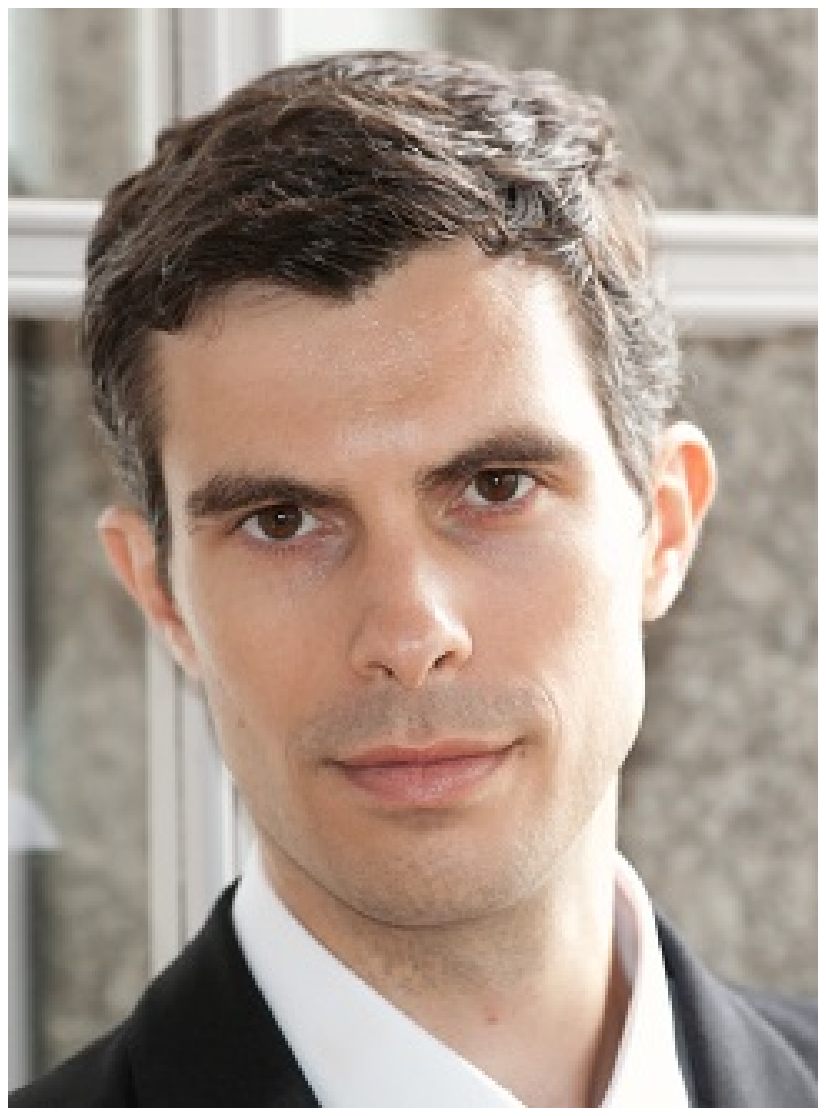}}]{Jos\'e P. Gonz\'alez Coma}
 was born in Mar\'in, Spain, in 1986. He received the Engineering Computer and M.Sc. degrees in 2009 and 2010 from Universidade da Coru\~na, Spain. Since 2009 he is with the Grupo de Tecnolog\'ia Electr\'onica y Comunicaciones (GTEC) at the Departamento de Electr\'onica y Sistemas. Currently, he is awarded with a FPI grant from Ministerio de Ciencia e Innovaci\'on. His main research interests are in designs of limited feedback and robust precoding techniques in MIMO wireless communication systems. Since Sep 2013 he is assistant professor in Universidade da Coru\~na.
\end{IEEEbiography}

\begin{IEEEbiography}[{\includegraphics[width=1in,height=1.25in,clip,keepaspectratio]{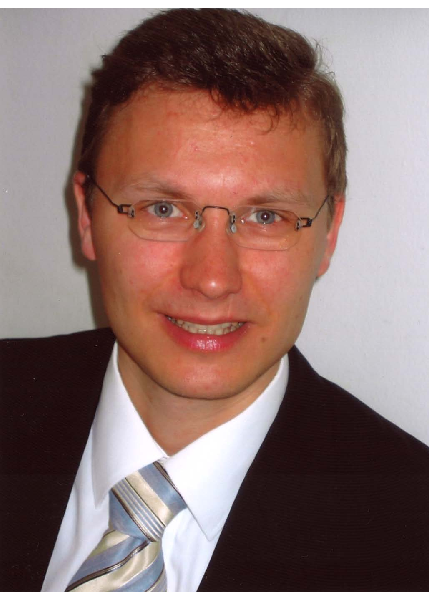}}]{Michael Joham}

was born in Kufstein, Austria, 1974. He received the Dipl.-Ing.
and Dr.-Ing. degrees (both summa cum laude) in electrical engineering
from the Technische Universit\"{a}t M\"{u}nchen (TUM), Germany,
in 1999 and 2004, respectively.

Dr. Joham was with the Institute for Circuit Theory and Signal Processing
at the TUM from 1999 to 2004. Since 2004, he has been with the Associate
Institute for Signal Processing at the TUM, where he is currently a senior
researcher. In the summers of 1998 and 2000, he visited the Purdue University,
IN. In spring 2007, he was a guest lecturer at the University of A coru\~na,
Spain. In spring 2008, he was a guest lecturer at the University of the German
Federal Armed Forces in Munich, Germany, and a guest professor at the
University of A coru\~na, Spain. In Winter 2009, he was a guest lecturer
at the University of Hanover, Germany. In Fall 2010 and 2011,
he was a guest lecturer at the Singapore Institute of Technology.
His current research interests are
precoding in mobile and satellite communications, limited rate feedback,
MIMO communications, and robust signal processing.

Dr. Joham received the VDE Preis for his diploma thesis in 1999 and the
Texas-Instruments-Preis for his dissertation in 2004. In 2007, he was a
co-recipient of the best paper award at the International ITG/IEEE Workshop
on Smart Antennas in Vienna. In 2011, he received the ITG Award 2011
of the German Information Technology Society (ITG).
\end{IEEEbiography}

\begin{IEEEbiography}[{\includegraphics[width=1in,height=1.25in,clip,keepaspectratio]{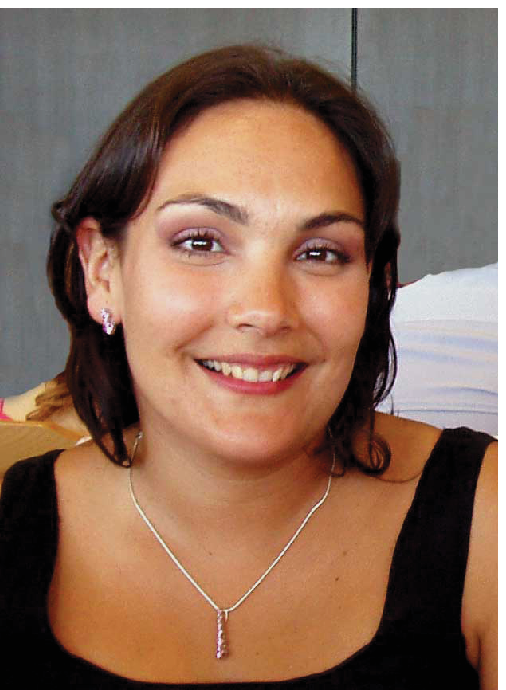}}]{Paula M. Castro}
 was born in Orense, Spain, 1975. She received her MS and PhD degrees in Electrical Engineering (2001) from the University of Vigo, Spain, and in Computer Engineering (2009) from the University of A Coru\~na, Spain, respectively. Since 2002 she is with the Department of Electronics and Systems at the University of A Coru\~na where she is currently an Associate Professor. Paula Castro is coauthor of more than forty papers in peer-reviewed international journals and conferences. She has also participated as a research member in more than thirty five research projects and contracts with the regional and national governments. She is a co-recipient of the Best Paper Award at the International ITG/IEEE Workshop on Smart Antennas, Vienna, 2007. Her research is currently devoted to the design of multiuser wireless communications systems.

\end{IEEEbiography}
\begin{IEEEbiography}[{\includegraphics[width=1in,height=1.25in,clip,keepaspectratio]{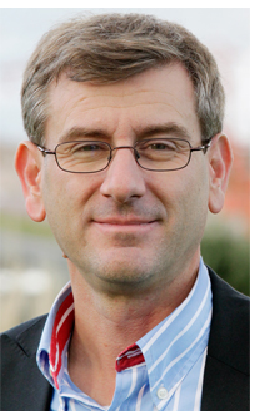}}]{Luis Castedo}
was born in Santiago de Compostela, Spain, in 1966.
He received the Ingeniero de Telecomunicaci\'on and Doctor
Ingeniero de Telecomunicaci\'on degrees, both from Universidad
Polit\'ecnica de Madrid (UPM), Spain, in 1990 and 1993, respectively.
Between November 1994 and October 2001, he has been an
Associate professor at the Departamento de Electr\'onica y Sistemas
at Universidad de A Coru\~na, Spain, where he is currently Full
Professor. He has been chairman of the Department between 2003
and 2009.

Luis Castedo is coauthor of more than one hundred and fifty papers
in peer-reviewed international journals and conferences. He has
also been principal researcher in more than thirty research projects
funded by public organisms and private companies. His research interests are signal processing and digital
communications with special emphasis on blind adaptive filtering,
estimation/equalization of MIMO channels, space-time coding and
prototyping of digital communication equipments.
\end{IEEEbiography}





\end{document}